# Adsorption-induced surface magnetism

*Miloš Baljozović*[†], *Shiladitya Karmakar*[‡], *André L. Fernandes Cauduro*[§,#], *Mothuku Shyam Sundar*[¶], *Marco Lozano*[‡], *Manish Kumar*[‡], *Diego Soler Polo*[‡], *Andreas K. Schmid*[§], *Ashutosh V. Bedekar*[¶], *Pavel Jelinek*[‡]* & *Karl-Heinz Ernst*[†,‡]*

[†]Empa, Swiss Federal Laboratories for Materials Science and Technology, Überlandstrasse 129, CH-8600 Dübendorf, Switzerland

[‡]Nanosurf Lab, Institute of Physics of the Czech Academy of Sciences, Cukrovarnická 10, Prague 6, CZ 162 00, Czech Republic

[§]National Center for Electron Microscopy, Molecular Foundry, Lawrence Berkeley National Laboratory, Berkeley, CA 94720, USA

[¶]Department of Chemistry, The Maharaja Sayajirao University of Baroda, Vadodara 390 002, India

e-mails: kalle@fzu.cz; jelinekp@fzu.cz

ABSTRACT

We report the emergence of adsorption-induced magnetism from heterohelicene molecules on a non-magnetic Cu(100) surface. Spin-polarized low-energy electron microscopy (SP-LEEM) measurements reveal spin-dependent electron reflectivity for enantiopure 7,12,17-trioxa[11]helicene (TO[11]H) monolayers, indicating the formation of a spin-polarized state localized in the topmost copper layer. Control experiments on clean Cu(100) and TO[11]H on highly oriented pyrolytic graphite show no such effect, excluding artifacts and chirality-induced spin selectivity as origins. Spin-polarized density functional theory calculations with hybrid functionals attribute the magnetism to strong chemisorption, which induces hybridization between the molecular HOMO and copper s- and d-states, driving asymmetric spin-polarized charge redistribution at the interface. An extended Newns–Anderson–Grimley model incorporating on-site Coulomb repulsion in Cu d-orbitals reproduces the emergence of interfacial spin polarization above a threshold interaction strength, highlighting the key roles of hybridization parameters and Coulomb correlation. These findings reveal a mechanism for inducing magnetism at molecule–metal interfaces without inherently magnetic components, offering avenues for engineering spin-polarized states in organic–inorganic hybrid systems.

## Introduction

In transition metals, paramagnetism originates from unpaired electrons in partially filled outer d-orbitals. These d-electrons are not only responsible for generating magnetic moments via their intrinsic spin and orbital angular momentum, but they also play a central role in the chemical bonding that governs the structure and properties of solid-state materials.[1] The relationship between covalent bonding and magnetism was first systematically explored by Linus Pauling, who laid the theoretical foundation for understanding magnetic exchange interactions through chemical bonding concepts.[2] Since then, both experimental and theoretical advances have significantly deepened our understanding of how electronic structure and bonding contribute to magnetic behavior in complex materials.[3-6]

A major challenge in molecular spintronics is the controlled manipulation of magnetism at the atomic and molecular scales—a requirement for next-generation spin-based logic and quantum technologies. One promising route relies on adsorption-induced magnetism, where non-magnetic systems acquire magnetic characteristics through charge transfer, orbital hybridization, or symmetry breaking at interfaces. Such effects have been observed for molecules on transition-metal surfaces, including the emergence of spin-polarized states at the $Cu/C_{60}$ interface.[7]

A growing body of work demonstrates that organic molecules can profoundly modify the magnetic properties of metallic substrates. π-Conjugated molecules can enhance exchange interactions and stabilize molecular-mediated magnetic units,[8] while $p_z$–d hybridization can even invert interfacial spin polarization and enable spin-selective electron injection.[9] The adsorption strength further dictates whether molecules harden or soften magnetic coupling, reflecting an interplay between local geometry and metal–molecule hybrid states.[10] Experiments on Cu-supported ferromagnets additionally show that organic acceptors can reverse magnetization and induce spin polarization via charge-transfer–driven π–d interactions.[11]

Complementary studies on self-assembled monolayers reveal that ordered organic interfaces on noble metals can exhibit significant magnetization,[12,13] high anisotropy, and spin-selective transport,[14] emphasizing the generality of molecularly driven magnetic phenomena. Together, these findings underscore the remarkable capacity of molecular adsorption to tailor interfacial magnetism in metallic systems. Despite this progress, the fundamental mechanisms governing molecular-induced magnetism in copper remain insufficiently understood, motivating a detailed investigation of how specific molecular interactions generate and control magnetic states at Cu surfaces.

Chiral organic molecules, such as helicenes, offer distinct advantages in the context of spin-dependent phenomena due to their intrinsic handedness and extended π-conjugated frameworks.[15,16] These structural features enable efficient electronic coupling with metal substrates and facilitate spin-selective interactions via the chirality-induced spin selectivity (CISS) effect.[17,18] While helicenes have been extensively studied for their optical and electronic properties,[19,20] their potential to induce or host magnetic moments upon adsorption on metal surfaces remains largely unexplored—despite detailed investigations into their adsorption behavior and self-assembly on various substrates.[21,22]

Recent work by Safari et al. demonstrated that heptahelicene exhibits partial enantioselective adsorption on Co islands supported on Cu(111), depending on the direction of out-of-plane magnetization.[23] Although the CISS effect is typically associated with spin-selective electron transport,[24-28] magneto-chiral selectivity during adsorption processes may also fall under this umbrella.[29] Together, such studies highlight the potential of hybrid organic–metal systems to exhibit emergent magnetic behavior, driven by interfacial charge redistribution and the ordered arrangement of chiral molecules at the surface.

Various techniques are available for studying surface magnetism, among which spin-polarized low-energy electron microscopy (SP-LEEM) is particularly well-suited for surface-sensitive investigations.[30,31] Owing to the short inelastic mean free path of ballistic electrons in solids,[32] SP-LEEM is highly surface sensitive, with the top few atomic layers of the sample primarily determining image contrast. In this technique, manipulation of the electron spin state is achieved by directing longitudinally spin-polarized electrons through an electromagnetic deflection system, followed by an electron-optical element. Together, these components allow precise orientation of the electron spin in any spatial direction (Figure 1a).

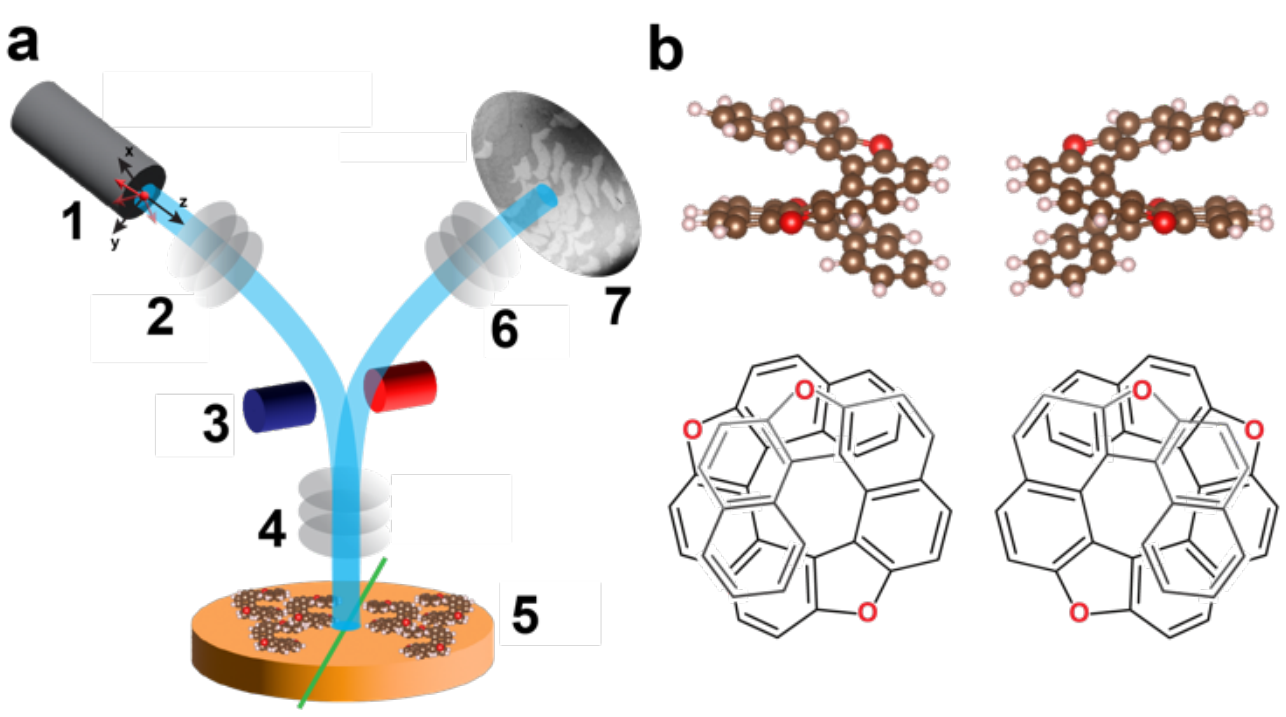

**Figure 1.** a) Sketch of SP-LEEM set-up, consisting of a spin-polarized electron gun with spin rotator (1), illumination optics (2), beam splitter (3), objective optics (4), the sample (5), imaging optics (6) and a detector/LEEM image (7). b) Structure of 7,12,17-trioxa[11]helicene (TO[11]H), shown as side views of ball-and-stick models (top) and skeletal formulas (bottom) for both enantiomers.

In this work, we report the emergence of adsorption-induced magnetism from heterohelicene molecules (Figure 1b) deposited on a Cu(100) surface. Spin-polarized low-energy electron microscopy (SP-LEEM) reveals that adsorption induces a spin-polarized state localized in the topmost atomic layer of copper. Density functional theory (DFT) calculations attribute this magnetism to strong chemisorption, which drives a complex charge backdonation mechanism at the interface, taking place between delocalized metallic s-bands and an antibonding state formed by interaction between the d-band and the HOMO orbital. Importantly, neither the molecular chirality nor direct charge transfer between the metal and the adsorbate contributes to the effect.

The origin of the magnetism is rationalized by employing an extended Newns–Anderson–Grimley model, which incorporates electron-electron interaction in spatially localized d-orbitals of the copper surface. The model reveals that strong hybridization of the molecular frontier HOMO orbital with surface states causes a complex charge backdonation between s- and d-bands, which—together with Coulombic repulsion in the localized d-orbitals—gives rise to spin polarization at the interface.

## RESULTS AND DISCUSSION

### Spin-polarized low-energy electron microscopy (SP-LEEM) studies

7,12,17-Trioxa[11]helicene (TO[11]H) molecules were adsorbed in their enantiopure forms—either *(P)*-TO[11]H or *(M)*-TO[11]H—onto the Cu(100) surface at room temperature. The formation of an ordered monolayer structure, previously observed by scanning tunneling microscopy (STM),[33,34] was monitored by low-energy electron microscopy (LEEM).[35]

Figure 2 displays electron reflectivity curves measured for a film of *(M)*-TO[11]H with slightly more than one close-packed monolayer. When the incident electron energy is below the sample's work function, all electrons are reflected. Upon reaching the work function threshold, reflectivity drops sharply, allowing precise determination of the work function from the reflectivity onset. Adsorption of around 1.3 ML of *(M)*-TO[11]H reduces the work function by approximately 1.0 eV (Figure 2a). Spatially resolved map of work function difference between 1$^{st}$ and 2$^{nd}$ molecular layer shows decrease in the work function of about 30 meV on the second layer (Figure S1).

Point-by-point measurements of spin asymmetry in electron reflectivity with respect to the kinetic energy of the incoming electrons have been performed across nearly the entire field of view (Figure S2). The superposition with the reflectivity curve shows that the spin-asymmetry occurs when electrons are reaching the interface (Figure 2b). Such spin-dependent reflectivity is a hallmark of a ferromagnetic surface, where the electronic structure and density of states differ for spin-up and spin-down electrons near the Fermi level. As in a ferromagnet, the exchange interaction causes energy splitting between majority and minority spin bands, electrons with spins aligned with the majority direction penetrate into the crystal, while those aligned with the minority direction are more likely to be reflected.[36] Because transmitted electrons are not detected in reflectivity measurements, higher reflectivity corresponds to lower available density of states. As a result, spin-up electrons are slightly more reflected than spin-down electrons, indicating a spin-selective interaction with the magnetic surface. Moreover, switching the electron polarization sequence leads as expected to the opposite effect, resulting in positive values of spin-asymmetries near the interface Figure S3.

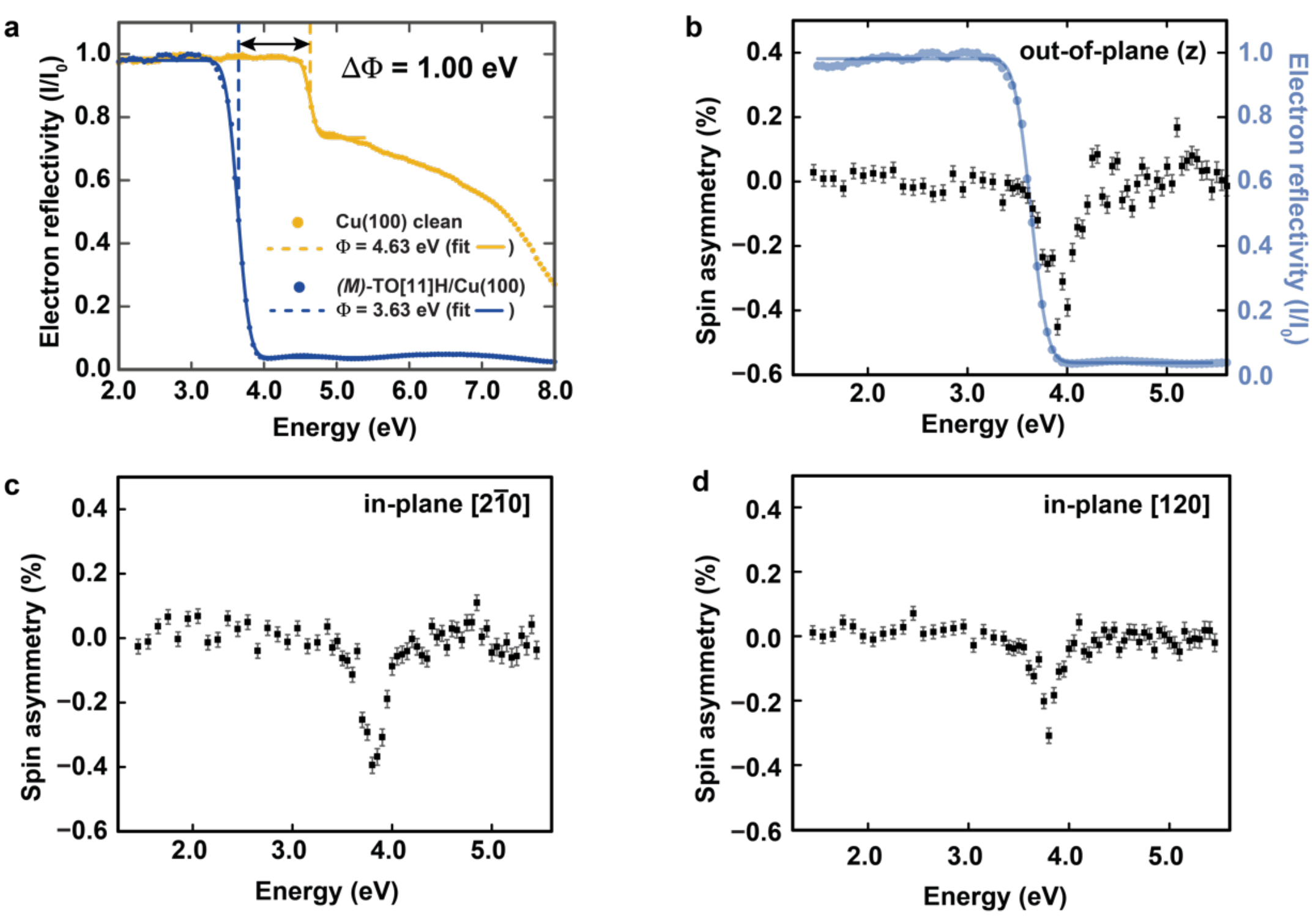

**Figure 2.** SP-LEEM-measured spin asymmetries. a) Electron reflectivity curves for clean Cu(100) (yellow) and of a 1.3 ML *(M)*-TO[11]H sample (blue). b) Points of spin asymmetry measurements and intensity of electron reflectivity (blue) with electron impact energy for out-of-plane spin alignment. c,d) Points of spin asymmetry measurements and intensity of electron reflectivity (blue) with electron impact energy for in-plane spin alignments along $[2\bar{1}0]$ and [120] directions.

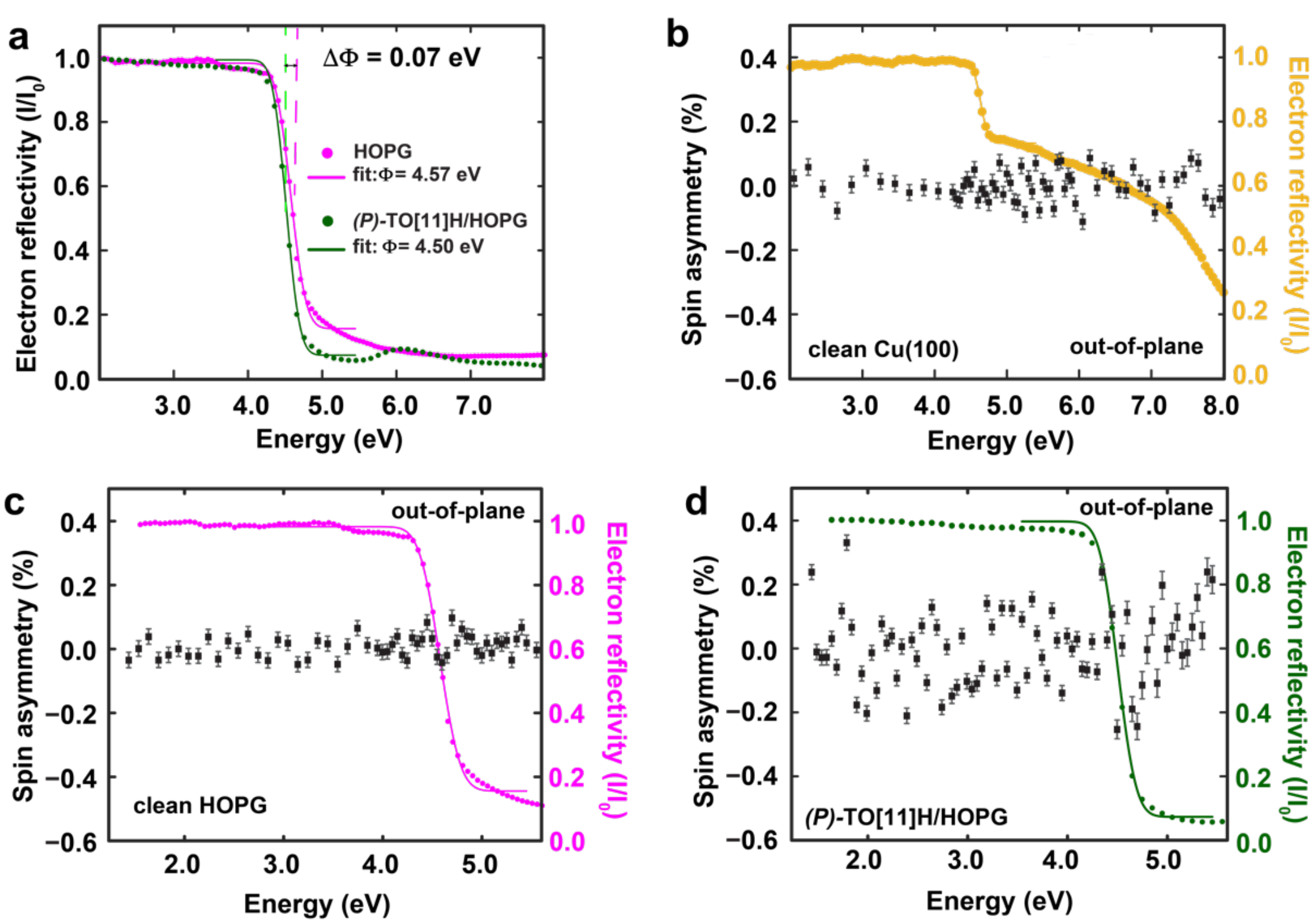

**Figure 3.** SP-LEEM-measured spin asymmetries for clean surfaces and *(P)*-TO[11]H on HOPG. a) Change of work function of HOPG due to adsorption *of (P)*-TO[11]H (1 ML). b,c,d) Spin asymmetry measurements on Cu(100), HOPG and a monolayer of *(P)*-TO[11]H on HOPG, respectively. For all three samples no asymmetries were observed.

Because the number of elastically backscattered electrons depends not only on surface structure but also on the relative orientation between the spin polarization of the incident beam and the surface magnetization, SP-LEEM allows probing of both in-plane and out-of-plane magnetization components by tuning the spin direction of the incoming electrons. Figures 2c and 2d display spin-asymmetry curves for two orthogonal in-plane spin orientations. Both curves show asymmetries comparable in magnitude to the out-of-plane case shown in Figure 2b. Although the observed spin asymmetries are relatively small, they are nonetheless significant. To verify that these signals are not artifacts or background effects, we compared them to spin asymmetry measurement of reference systems where no magnetism is expected. As shown in Figure 3, clean Cu(100) and TO[11]H molecules adsorbed on highly oriented pyrolytic graphite (HOPG) exhibit no measurable spin asymmetry within the experimental error.

Since the observed effect is identical in sign for both enantiomers (Figure S4), a contribution from the CISS effect to electron reflectivity can be reliably excluded. Previous SP-LEEM studies on heptahelicene adsorbed on Cu(100) also reported no evidence of CISS,[37] suggesting that the origin of the effect may be more closely linked to topological modifications in the metallic substrate induced by the chiral molecules, rather than arising from the molecular states alone. It is worth noting, however, that spin-resolved photoelectron spectroscopy studies conducted in vacuum on similar systems have produced contradictory results, highlighting the complexity and sensitivity of CISS-related phenomena to experimental conditions.[38,39] Similar conclusions have been made in a recent study based on break junction experiments and theoretical considerations.[40]

### Results of DFT Calculation

From the optimized model structure, bond formation between Cu atoms and four C atoms of the proximal C6 ring of the TO[11]H molecule is observed. As the bond lengths are only between 2.163 Å and 2.251 Å, a strong chemisorption of the TO[11]H molecule on the Cu(100) surface is concluded. The binding energy of the model system is calculated to be –4.33 eV, revealing a very strong binding of the TO[11]H molecule on the Cu(100) surface. Such value is similar to the reported binding of curved, bowl-shaped polycyclic aromatic hydrocarbons (so-called buckybowls) on copper.[41,42] The calculated charge density difference ($\Delta\rho$) provides further evidence of strong chemisorption of the TO[11]H molecule on the Cu(100) surface. As shown in Figure S5, significant charge accumulation is observed around the carbon atoms of the six-membered ring in the TO[11]H molecule, accompanied by charge depletion around the underlying bonded Cu atoms. Consequently, this further confirms the robust chemisorption.

The spin-polarized DFT calculations with Perdew-Burke-Ernzerhof (PBE) functional results in non-magnetic ground state with symmetric spin density of states. This suggests that PBE, which tends to underestimate electronic correlation, is insufficient to capture the magnetic instability at the interface. To more accurately describe the strongly correlated electronic interactions necessary for magnetism, we employed the hybrid PBE0 functional, which incorporates a portion of exact Hartree-Fock exchange. On a reduced model system (Figure 4a), the PBE0 functional provide a spin-polarized ground state. The computed spin density (isosurface value = 0.02 e/Å$^3$, Figure 4b) is primarily localized on the carbon atoms of the chemisorbed ring, with a weaker but non-negligible contribution on the directly bonded Cu atoms beneath.

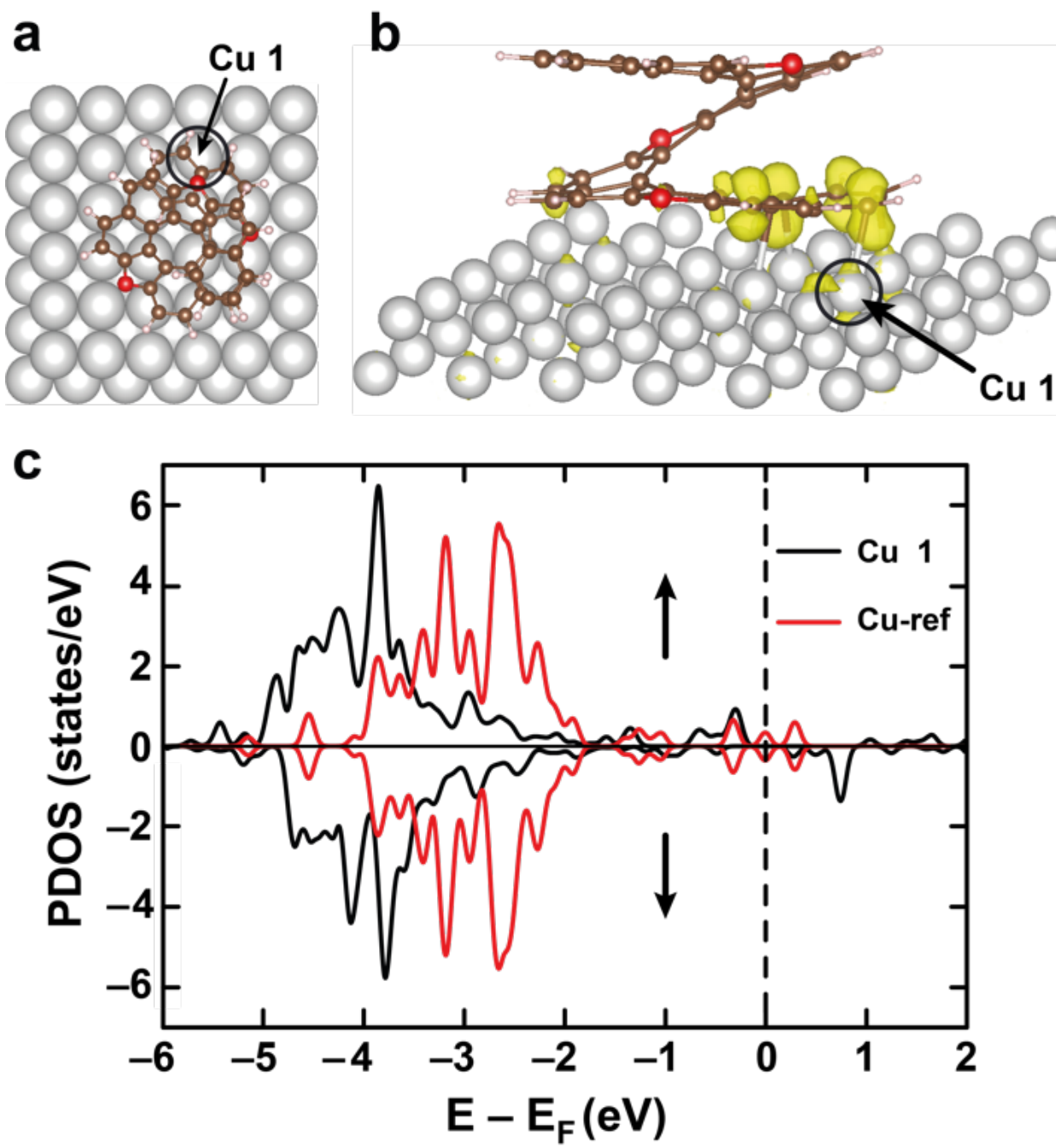

**Figure 4.** a) Ball-and-stick representation (top view) of the model system. Grey: Cu, brown: C, red: O, pink: H. b) Side view of the model system along with the plot of spin density accumulation at the interface (isosurface maximum value set at 0.02 e/Å$^3$). c) Density of states projected to Cu1 attached to TO[11]H and to a Cu atom (Cu-ref) on a clean Cu(100) surface. (Cu1 marked by arrows and circles in **a** and **b**).

The absence of spin polarization in PBE, contrasted with its presence in PBE0, suggests that the inclusion of exact exchange in PBE0 is essential for stabilizing the spin-polarized state. To systematically assess the role of exact exchange, the contribution of nonlocal Hartree–Fock exchange ($\varepsilon_{xc}$) was varied from 5% to 35%. As shown in Figure S6, no spin density appears at 5% $\varepsilon_{xc}$, but begins to emerge around 15% and grows progressively stronger with increasing $\varepsilon_{xc}$. Note that the standard PBE0 functional corresponds to $\varepsilon_{xc}$ = 25%.

Figure 4c presents the projected density of states (PDOS) for two copper environments: Cu1, which is directly bonded to the carbon atoms of the TO[11]H molecule's six-membered ring, and a reference Cu atom from a clean Cu(100) surface. Near the Fermi level, the PDOS is dominated by s-orbital contributions, while the d-orbitals are centered around –3.8eV for Cu1 and –2.5eV for the reference Cu atom. The low-energy position of the Cu1 d-band reflects the strong bonding interaction with the TO[11]H molecule, whereas the reference Cu atom presents an electronic structure characteristic of the bare surface.

Moreover, the reference Cu atom exhibits a symmetric, non-spin-polarized PDOS. In contrast, Cu1 displays a pronounced asymmetry in the PDOS below –4 eV, corresponding to its d-orbital region. Note that the apparent suppression of the Cu1 DOS at the Fermi level does not indicate insulating behavior. It arises from the finite slab thickness (only 2 layers) and the use of Γ-point sampling, which was necessary because hybrid exchange-correlation functional calculations for metallic slabs are computationally very demanding.

While DFT simulations with hybrid functionals clearly demonstrate that magnetism originates from the hybridization of d-electrons with the HOMO orbital and from enhanced electron–electron interactions in localized states, they do not readily reveal the exact mechanism responsible for interfacial magnetism. To address this limitation, we analyzed model Hamiltonians to gain deeper insight into the magnetic behavior at the molecule/metal interface.

### Results of Model Hamiltonian Calculation

One of the standard models to describe the mechanism of an adsorbate bonding on a metal surface is the Newns–Anderson–Grimley model,[43,44] which provides insights about the hybridization between the adsorbate molecule and the electronic band structure of the metal surface. According to this model, hybridization with broad metal valence bands, such as s-bands, leads to a broadened adsorbate state. In contrast, interaction with narrow valence bands, as d-bands, can result in the splitting of the adsorbate state into localized bonding and antibonding states.

To gain deeper insight into the origin of interface magnetization, the Newns–Anderson–Grimley model was extended by incorporating on-site Coulomb electron–electron interactions in both the HOMO ($U_{HOMO}$) and the localized d-band ($U_d$). A detailed description of the model is provided in the Methods section and illustrated in Figure 5a. The results demonstrate that the magnitude of the Coulomb on-site interaction in the d-orbital plays a key role in the emergence of spin polarization at the interface.

The model parameters were chosen such that they reproduce the qualitative electronic structure of Cu surface and the molecule, rather than to fit experimental data. To mimic the electronic structure of the Cu(100) surface, delocalized Cu-s states were represented by a broad band centered at the Fermi level with on-site energy of the s-orbitals to $\varepsilon_s = 0$ eV and hopping parameter T = 1.5 eV to reflect the broad Cu-s bandwidth. One electron per $\varepsilon_s$-site was assumed to create a half-occupied metallic s-band. Additionally, a doubly occupied Cu-d band was approximated by a localized d-state located $\varepsilon_d \approx -2$ eV below the Fermi level. The on-site interaction $U_d$ was scanned from 0 to 4 eV, consistent with reported values for Cu-3d states.[45] For simplicity, hopping between d-orbitals was neglected to approximate the narrow-band limit.

According to DFT calculations, the molecular HOMO remains doubly occupied and appears at $\varepsilon_{HOMO} = -0.5$ eV. The model Hamiltonian was solved using the self-consistent one-electron mean-field approximation, see Methods. The validity of the one-electron mean-field solution was confirmed by density matrix renormalization group (DMRG) calculations solving the model Hamiltonian.

Figure 5b shows the reference PDOS projected onto the HOMO, frontier s-site, and d-site in the absence of molecule–surface coupling ($t_{sh}$ = $t_{dh}$ = 0 eV) and Coulomb interactions ($U_d$ = 0 eV). The singly occupied s-band appears broadened around the Fermi level, while the doubly occupied d-site and HOMO remain localized near –2 eV and –0.5 eV, respectively. Introducing the molecule–surface interaction via $t_{sH}$ and $t_{dH}$ leads to hybridization of the HOMO with both the s-band and d-state.

In the absence of the coupling between HOMO and s-band, $t_{sH}$ = 0, hybridization between the HOMO and the d-state ($t_{dh}$ = 1 eV) produces bonding and antibonding states, both of which are initially doubly occupied. The magnitude of the on-site Coulomb interaction of d-state, $U_d$, sets the position of the antibonding state. Figure 5c shows the calculated DOS for $U_d$ = 4 eV and $t_{sH}$ = 0, where the antibonding state is located above the Fermi level set by the center of the s-band. This results in the net charge transfer from the antibonding state towards the s-band in the mean-field solution. Consequently, the antibonding state becomes completely empty, and two electrons are uniformly distributed into the s-band. Nevertheless, the net charge transfer does not cause the magnetization at the interface, as demonstrated by the symmetric DOS in both spin channels shown in Figure 5c.

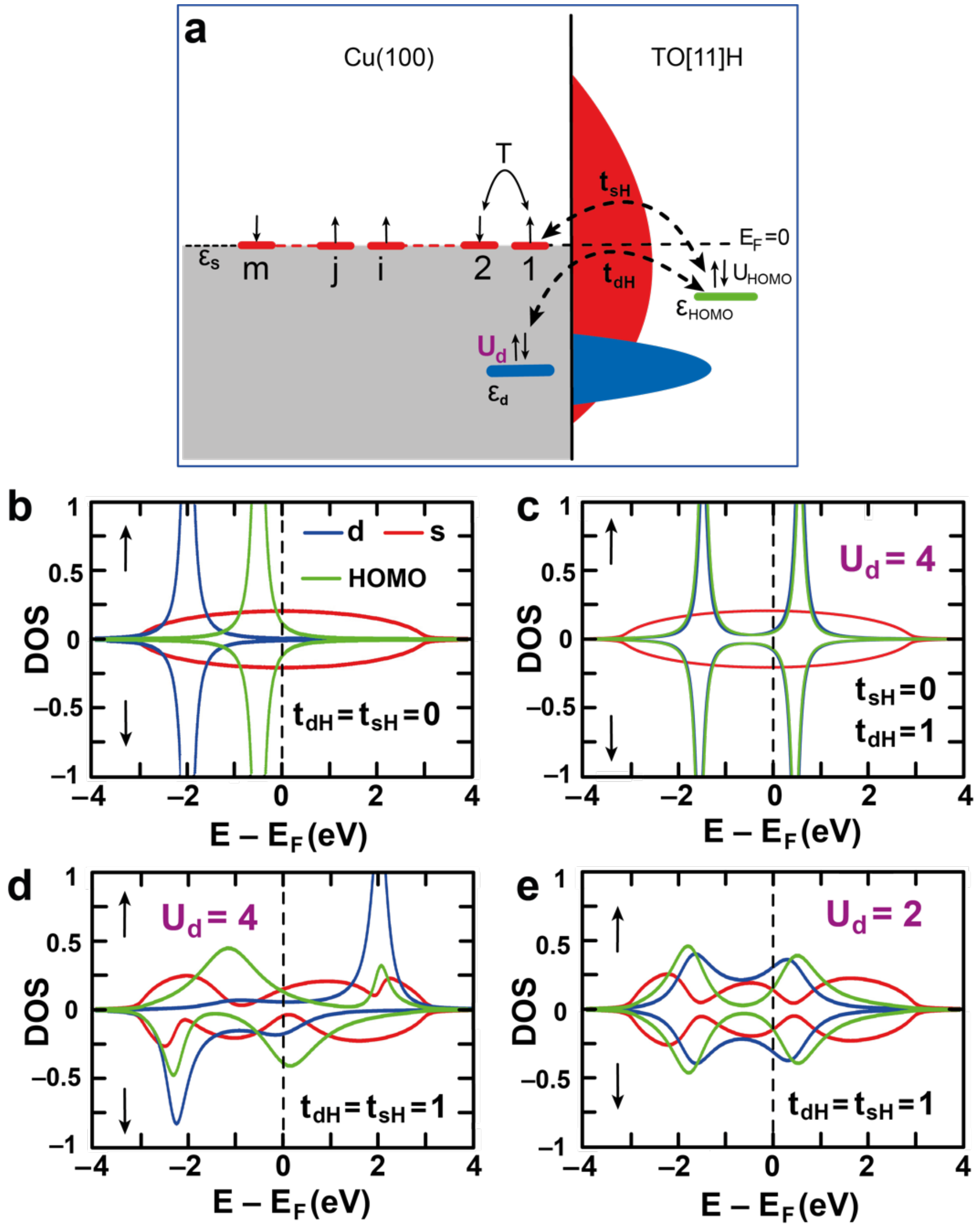

**Figure 5.** (a) Schematic representation of the model Hamiltonian calculation. The Fermi level, $E_F = 0$, is marked with a black dashed line, with the grey-shaded region below $E_F$ showing occupied substrate bands. Half-filled Cu-s chain (red lines), fully occupied Cu-d (blue) and TO[11]H-HOMO (green) are shown, with shaded red and blue indicating delocalized s and localized d states. Spin-up/spin-down electrons are shown by up/down arrows. s-chain hopping T and coupling interactions ($t_{sH}$, $t_{dH}$) are shown by solid and dashed double-headed arrows respectively. Projected DOS onto TO[11]H-HOMO, Cu-d, and frontier Cu-s sites shown for: (b) $U_d = 0$, $t_{sH} = t_{dH} = 0$; (c)

$U_d$ = 4 eV, $t_{sh}$ = 0, $t_{dh}$ = 1 eV; (d) $U_d$ = 4 eV; (e) $U_d$ = 2 eV, $t_{sh}$ = $t_{dh}$ = 1 eV. Model parameter used: $\varepsilon_s$ = 0 eV, m = 100, $\varepsilon_d$ = −2eV, $\varepsilon_{HOMO}$ = −0.5eV, T = 1.5 eV, $U_d$ = 0 to 4 eV (as indicated), $U_{HOMO}$ = 0.

Interestingly, turning on the coupling between HOMO and s-band, $t_{sH}$ = 1 eV, causes the interface magnetism manifested by strongly asymmetric DOS between spin up and spin down channels, as shown in Figure 5d. A strong modification of both bonding and anti-bonding states originally formed by localized d- and HOMO states is observed. Namely, a strong split of d-state in spin channels occurs, where d-state in the spin-up channel is localized at ~ 2 eV above the Fermi level, therefore becoming completely empty. On the contrary, the d-state in the spin-down channel is located ~ 2 eV below the Fermi level occupied by one electron. Similarly, the HOMO state also becomes asymmetric in both spin channels. The resulting occupations are $n_{d,\uparrow}$ = 0.12, $n_{d,\downarrow}$ = 0.89; $n_{HOMO,\uparrow}$ = 0.76, $n_{HOMO,\downarrow}$ = 0.55; and $n_{s,\uparrow}$ = 0.46, $n_{s,\downarrow}$ = 0.51. These values clearly reflect the spin imbalance, with the d-orbital showing a strong preference for spin-down electrons, while the HOMO favors spin-up electrons. These results point out that the coupling between the delocalized s-band and HOMO orbital causes asymmetric charge backdonation in two spin channels, facilitating the emergence of the interface magnetization.

But the emergence of the interface magnetization also strongly depends on the on-site Coulomb interaction at the d orbital. Figure 5e displays the calculated DOS for the case, where the on-site Coulomb interaction at the d-state is set to $U_d$ = 2 eV. Notably, spin up and spin down occupations remain equal, resulting in no spin polarization, as seen from the symmetric DOS of the hybridized states in Figure 5e. According to the simulation, the HOMO loses more charge than the d orbital, and the excess charge in the frontier s orbital is redistributed uniformly across the remaining s orbitals in the tight-binding chain.

It is instructive to analyse how different parameters of the model Hamiltonian affect the emergence of the interfacial magnetism. According to the model, the parameters directly influencing spin polarization in the molecule are: 1) hopping parameters $t_{sH}$ and $t_{dH}$, which determine the degree of orbital hybridization between the molecule and the metal. 2) The on-site Coulomb interaction at the d-orbital of Cu, *i.e.*, $U_d$. For fixed hopping parameters $t_{sH}$ and $t_{dH}$, spin polarization is induced at HOMO once $U_d$ exceeds a certain threshold ($U_{d\text{-threshold}}$). Figure S7 provides phase-diagrams displaying the magnetic and non-magnetic regions in the ($U_d$, $t_{dH}$) and ($U_d$, $t_{sH}$) parameter spaces.

The onsite energies $\varepsilon_s$ and $\varepsilon_d$ are fixed for a given metal surface, while the hopping parameter between the metal's s-orbitals depends on the bandwidth of the metal. In contrast, the values of $\varepsilon_{HOMO}$ and $U_{HOMO}$ are molecule-specific. Our model allows to explore the conditions under which spin polarization emerges by systematically varying $\varepsilon_{HOMO}$ and $U_{HOMO}$. To visualize the influence of these parameters, a phase space was constructed involving $\varepsilon_{HOMO}$ and $U_{HOMO}$ and $U_d$, illustrating how a magnetic moment can develop at the HOMO site as shown in Figure S8. For simplicity, the hopping parameters $t_{sH}$ and $t_{dH}$ were fixed at 1. The phase diagram in Figure S8 illustrate the non-magnetic and magnetic regions, with magnetization strength determined by $M_{HOMO}$. While the $U_{HOMO}$ - $U_d$ diagram shows a sharp magnetic transition, the $\varepsilon_{HOMO}$ - $U_d$ diagram displays more gradual boundary. The emergence of magnetism at the interface is also confirmed by many-body DMRG calculations resolving the model Hamiltonian for different parameters of $U_{HOMO}$ and $U_d$, see Figure S9.

In contrast to the modified Stoner mechanism that has been proposed for chemisorbed systems such as $Cu/C_{60}$,[Error! Bookmark not defined.] the spin polarization in the present model is primarily governed by three factors: (i) the HOMO–Cu hybridization terms ($t_{sH}$, $t_{dH}$, Figure S7), (ii) the on-site interaction on Cu-d ($U_d$, Figure S7), and (iii) the alignment of the HOMO level ($\varepsilon_{HOMO}$) relative to the Fermi energy (Figure S8b). Because magnetism arises from the interplay of these parameters, a simple rule of thumb is not readily available. This behaviour is more naturally captured within an Anderson-type framework, where the key control parameter is the ratio U/t. Consequently, the origin of interfacial magnetism extends beyond the single-band Stoner picture based on the product of density of states at the Fermi level and Coulomb interaction, $D(E_F)U$.

Nonetheless, our analysis reveals clear qualitative trends. The phase diagrams in Figures S7 and S8 delineate magnetic and non-magnetic regions and show a threshold value of $U_d \approx 2$ eV for the onset of magnetism. We further find that this threshold depends only weakly on $U_{HOMO}$, and $t_{sH}$. In contrast, Figure S7a shows that large $t_{dH}$ suppresses the magnetic solution. The magnetic phase boundary in the $t_{dH}$-$U_d$ plane is essentially linear, with a slope of ~ 0.3, indicating that magnetism appears only when $U_d/t_{dH} \approx 3.33$. Increasing $t_{dH}$ delocalizes the HOMO and consequently shifts the onset of magnetism to larger $U_d$, reflecting the interplay between hybridization and Coulomb correlation. Finally, the HOMO level must lie close to the Fermi energy, but not more than ~ 1.5 eV below it.

## Conclusions

Combined SP-LEEM experiments, spin-polarized DFT calculations, and extended Newns–Anderson–Grimley modeling demonstrate that strong chemisorption of heterohelicene molecules on Cu(100) can induce interfacial spin polarization in an otherwise non-magnetic substrate. The effect arises from hybridization between the molecular HOMO and copper s- and d-states, with Coulomb interactions in the localized d-orbitals driving a spin-symmetry breaking above a critical threshold. Notably, the magnetism is independent of molecular chirality and does not require direct charge transfer from the adsorbate to the substrate. These findings establish adsorption-induced magnetism as a robust route for engineering spin-polarized states in organic–metal interfaces, expanding the design space for molecular spintronic architectures that do not rely on intrinsically magnetic components.

## METHODS

### Experimental

TO[11]H has been synthesized as reported previously.[46] Enantiomer separation via high-performance liquid chromatography, their circular dichroism and fluorescence is provided in the Supporting Information (Figures S10-S12). The Cu(100) surface has been cleaned by repetitive argon ion sputtering and annealing at 830 K. In the first several cycles of sputtering, small amount of oxygen was added for carbon removal. The cleanliness of the substrate was confirmed by Auger spectroscopy. *(M)*- and *(P)*-TO[11]H molecules were deposited on substrates kept at room temperature from home-made effusion cells held at 480 K. Deposition was performed directly in the SP-LEEM chamber and coverage was monitored in situ by LEEM intensity/reflectivity changes.

The SP-LEEM measurements were performed at the National Center for Electron Microscopy of the Lawrence Berkeley National Laboratory. Cesiated GaAs cathode with the peak of energy distribution of emitted electrons at around $E_C^0$ = 1.4 eV was used as an electron source. All samples were prepared under ultrahigh vacuum (UHV) conditions directly in the SP-LEEM chamber, with a base pressure lower than $2.0 \cdot 10^{-10}$ mbar. Samples were initially corrected for the tilt and electron beam aligned for homogeneous irradiation over the complete field of view with out-of-plane polarized electron spins. In order to avoid beam damage, the measurements were continuously performed on different areas by tediously moving the electron beam laterally over the sample. At each energy, detector gain was adjusted to compensate for the reflected beam intensity drop due to reflectivity change. The spin-asymmetry of the probed area is defined *via* the Sherman function as:

$$A\ (\%) = \frac{R_{\uparrow} - R_{\downarrow}}{R_{\downarrow} + R_{\uparrow}} \cdot 100 \qquad (1),$$

with $R_{\uparrow}$ and $R_{\downarrow}$ being the reflectivity of spin up and down electrons and P the polarization of the beam. Error bars in the plots represent Standard Error of the Mean obtained from the repeated measurements at the same beam energy.

## Computational Details

### DFT calculations

To theoretically investigate the TO[11]H/Cu(100) interface, we constructed a model system to simulate the interaction between the TO[11]H molecule and the Cu(100) surface. In this model, the TO[11]H molecule was placed on top of two atomic layers of the Cu(100) surface, with each layer containing 10 Cu atoms arranged periodically along the *a*- and *b*-directions within a simulation cell. The in-plane lattice parameter of the simulation cell is as follows: a = b = 25.6 Å, which contains a total of 263 atoms. To minimize the artificial interaction between the TO[11]H/Cu(100) composite and its images along the out-of-plane direction within the periodic setup of the calculation, a vacuum space of ~28 Å is used. Ab initio study on the model system is performed within the framework of DFT using the all-electron electronic structure code FHI-aims with the 'light' default settings for the numerical grid and basis set.[47] Structural optimization of the model structure is carried out using the exchange-correlation approximated through the Perdew-Burke-Ernzerhof formulation of generalized gradient approximation (PBE-GGA).[48] Dispersive forces were included in the calculations using the Tkatchenko-Scheffler (TS) method,[49] while relativistic effects were treated on the level of the atomic zero-order regular approximation (atomic ZORA).[47] The self-consistent field (SCF) cycle was treated as converged when changes of total energy, sum of eigenvalues and charge density were found below $10^{-6}$ eV, $10^{-3}$ eV and $10^{-5}$ eÅ$^{-3}$, respectively. During the structural optimization, the bottom Cu layer was constrained, and only the ionic positions of TO[11]H and the top Cu layer were relaxed within the simulation cell until the Hellmann–Feynman forces were reduced below 0.005 eVÅ$^{-1}$. The Brillouin zone was sampled using the Gamma point. To obtain the PDOS of a reference Cu atom, we performed PBE0 calculation for the reduced model system after removing the TO[11]H molecule, *i.e.*, for the clean Cu(100) surface.

For stability analysis of the TO[11]H/Cu(100) interface, the binding energy was evaluated by: $E_{binding} = E_{model} - E_{TO[11]H} - E_{Cu(100)}$, where $E_{model}$ denotes the total energy of the model system containing the TO[11]H molecule located on Cu(100) surface. $E_{TO[11]H}$ and $E_{Cu(100)}$ denote the total energy of the TO[11]H molecule and the Cu(100) surface, respectively. To determine any potential charge transfer between TO[11]H and the Cu(100) surface, the charge density difference is calculated via: $\Delta\rho = \rho_{model} - \rho_{TO[11]H} - \rho_{Cu(100)}$, Whereby $\rho_{model}$, $\rho_{TO[11]H}$ and $\rho_{Cu(100)}$ denote the corresponding charge densities of the model, TO[11]H molecule and the Cu(100) surface, respectively.

To investigate magnetism at the TO[11]H/Cu(100) interface, we have employed spin-polarized DFT calculations within PBE as well as the hybrid functional PBE0 schemes,[50] modified by changing the portion of the Fock exchange ($\varepsilon_{xc}$) contributing to this functional. $\varepsilon_{xc}$ is varied from 5% to standard 25% to 35% of non-local Fock exchange. To make the calculation tractable within the PBE0 scheme, the system size was reduced from a 10×10 to a 6×6 supercell. This scaling reduces the number of atoms but does not alter the underlying lattice periodicity in the in-plane directions. Note that the bonding environment around the TO[11]H/Cu(100) interface remains unaltered by this cell reduction. The in-plane lattice parameter of the reduced simulation cell becomes as follows: a = b = 15.3 Å, containing a total of 135 atoms, thus reducing the number of atoms inside the cell and making it suitable for hybrid calculations. Converged SCF is used to compute the spin density on the reduced model system within both PBE and PBE0 schemes.

**Model Hamiltonian**

To understand the influence of electron-electron interactions and hybridization on the electronic and magnetic properties of the TO[11]H/Cu(100) interface, a Hubbard model expressed by following Hamiltonian has been employed:

$$H = H_{HOMO} + H_s + H_d + H_{sH} + H_{dH} \quad (2).$$

The first term in Eq. 2 for the adsorbate TO[11]H molecule, arising from its frontier HOMO orbital, can be expressed as:

$$H_{HOMO} = \sum_{\sigma} \left( \varepsilon_{HOMO} C^{\dagger}_{HOMO\sigma} C_{HOMO\sigma} \right) + U_{HOMO} n_{HOMO\uparrow} n_{HOMO\downarrow} \quad (3).$$

It consists of two parts, i) the on-site energy of electrons occupying HOMO, and ii) Coulomb repulsion from double occupancy of HOMO. Here $\varepsilon_{HOMO}$ and $U_{HOMO}$ represent energy level and on-site Coulomb interaction of HOMO respectively. The operator $C^{\dagger}_{\sigma}$ ($C_{\sigma}$) creates (annihilates) an electron with spin σ on any state. $n_\uparrow$ and $n_\downarrow$ are the number operators for electrons with spin-up and spin-down, respectively.

The second and third terms in Eq. 2 represent the Cu(100) substrate. The second term arises from the delocalized 4s states, which are singly occupied and contribute to conduction. The third term originates from the localized 3d states, which are fully occupied and do not participate in conduction. However, they can hybridize with the TO[11]H molecule and, under certain electron correlation, may induce spin polarization, as suggested by DFT results. For modeling conduction 4s electrons of Cu, we consider a simple tight-binding model of linear chain of site i (i=1,2,.,m; m=100) characterized by onsite energy $\varepsilon_s$ and nearest neighbor hopping integral T. Hence the second term can be expressed as:

$$H_s = \sum_{i,\sigma} \varepsilon_s C_{i,\sigma}^\dagger C_{i,\sigma} + T \sum_{ij>,\sigma} \left( C_{i,\sigma}^\dagger C_{j,\sigma} + h.c. \right) \quad (4).$$

The third term corresponding to the localized 3d electrons can be expressed as:

$$H_d = \sum_\sigma \left( \varepsilon_d C_{d\sigma}^\dagger C_{d\sigma} \right) + U_d n_{d\uparrow} n_{d\downarrow} \quad (5).$$

Here $\varepsilon_d$ and $U_d$ represent energy level and on-site Coulomb interaction of d-orbitals respectively.

The fourth and fifth terms in Eq.1 correspond to the coupling of TO[11]H adsorbate on Cu(100) substrate, representing hybridization of Cu-4s and Cu-3d orbitals with TO[11]H-HOMO respectively and can be expressed as:

$$H_{dH} = t_{dH} \sum_\sigma \left( C_{d\sigma}^\dagger C_{H\sigma} + h.c. \right) \quad (6),$$

$$H_{sH} = t_{sH} \sum_\sigma \left( C_{s\sigma}^\dagger C_{H\sigma} + h.c. \right) \quad (7).$$

Here $t_{dH}$ and $t_{sH}$ represent the corresponding hopping integral.

### Mean Field Approximation

The Hubbard model Hamiltonian was solved using the Hartree-Fock mean-field approximation under the assumption of collinear spins. Under this approximation, the many-body Coulomb interaction term can be simplified as follows:

$$Un_{i,\uparrow}n_{i,\downarrow} = Un_{i,\uparrow}\langle n_{i,\downarrow}\rangle + Un_{i,\downarrow}\langle n_{i,\uparrow}\rangle - U\langle n_{i,\uparrow}\rangle\langle n_{i,\downarrow}\rangle \quad (8),$$

where $\langle n_{i,\sigma}\rangle$ ($\sigma = \uparrow, \downarrow$) is the average or "mean-field" spin σ electron density at site i. Hence the problem reduces to an effective single-electron problem characterized by Mean Field Hubbard (MFH) Hamiltonian, where each electron experiences a mean-field potential that depends on the electron density. The electron density is determined self-consistently from the single-electron eigenstates. A random electron density $\langle n^0_{i,\sigma}\rangle$ was used as a starting guess. Keeping the mean-field electron density constant, the MFH Hamiltonian was divided into spin up and down terms, and solved for its single particle eigenvalues $E_{j,\sigma}$ and eigenvectors $C_{i,j,\sigma}$ (i: site index, j: eigenstate index). Therefore, the electron density can be calculated as:

$$\langle n_{i,\sigma}\rangle = \sum_j^{N_{el}} \left|C_{i,j,\sigma}\right|^2 \quad (9).$$

The difference between the input and output electron densities was quantified by the residual:

$$Y = \sqrt{\sum_i \left(\langle n_{i,\sigma}\rangle - \langle n^0_{i,\sigma}\rangle\right)^2} \quad (10).$$

If the residual was sufficiently small, then the calculation reached self-consistency. Otherwise, the computed $\langle n_{i,\sigma}\rangle$ values will be used to update charge densities ($\langle n^0_{i,\sigma}\rangle$) via the linear mixing algorithm:

$$\langle n^0_{i,\sigma}\rangle = \beta\langle n^0_{i,\sigma}\rangle + (1-\beta)\langle n_{i,\sigma}\rangle \quad (11),$$

here, β is mixing parameter lying between 0 and 1.

The parameters of interest are the following,

1) The magnetic moments on each site i:

$$M_i = \langle n^0_{i,\uparrow} \rangle - \langle n^0_{i,\downarrow} \rangle \quad (12).$$

2) The density of states on each site i for spin σ:

$$DOS_{i,\sigma}(E) = \eta \sum_j \frac{|c_{i,j,\sigma}|^2}{(E - E_{j,\sigma})^2 + \eta^2} \quad (13),$$

where, η is the broadening factor.

**DMRG Calculations**

The density matrix renormalization group (DMRG) method was employed to obtain the ground-state properties of the model Hamiltonian. All DMRG calculations were performed using the ITensor library in Julia.[51] A truncation error cutoff of $10^{-8}$ was imposed to ensure high numerical accuracy, while the bond dimension was allowed to grow without an explicit upper bound. The variational optimization of the matrix product state was iterated until the total energy converged within $10^{-8}$ eV.

AUTHOR INFORMATION

**Corresponding Authors**

Karl-Heinz Ernst; orcid 0000-0002-2077-4922; kalle@fzu.cz;

Pavel Jelinek; orcid 0000-0002-5645-8542; jelinekp@fzu.cz

**Present Address**

# Nova measuring instruments, Fremont, CA, 94538, USA

**Notes**

The authors declare no competing financial interest.

ACKNOWLEDGMENT

Work at the Molecular Foundry was supported by the Office of Science, Office of Basic Energy Sciences, of the U.S. Department of Energy under Contract No. DE-AC02-05CH11231. Support by the Swiss National Science Foundation (grant number 212167), the Grantová Agentura České Republiky - GAČR (project 24-11064S), and University of Zurich Research Priority Program LightChEC is gratefully acknowledged.

ASSOCIATED CONTENT

**Supporting Information**.

The Supporting Information is available free of charge at ...

LEEM images, work function difference maps, spin asymmetry histogram and image, inverted spin-polarization plot, spin asymmetry plot for *(P)*-TO[11]H, charge density difference and spin density plots of adsorbate complex, phase diagrams showing dependence of magnetization on various modeling parameters, details of HPLC separation (PDF).

REFERENCES

(1) Goodenough, J. B. Magnetism and the Chemical Bond; John Wiley & Sons: Hoboken, NJ, USA, 1963.

(2) Pauling, L. The nature of the chemical bond. Application of results obtained from the quantum mechanics and from a theory of paramagnetic susceptibility to the structure of molecules. *J. Am. Chem. Soc.* **1931**, *53*, 1367–1400.

(3) Hubbard, J.; Marshall, W. Covalency effects in neutron diffraction from ferromagnetic and antiferromagnetic salts. *Proc. Phys. Soc.* **1965**, *86*, 561–572.

(4) Tofield, B. C. Covalency effects in magnetic interactions. *J. Phys. Colloques* **1976**, *37*, 539–570.

(5) Anderson, P. W. Theory of Magnetic Exchange Interactions: Exchange in Insulators and Semiconductors. In Solid State Physics; Seitz, F., Turnbull, D., Eds.; Academic Press: Cambridge, MA, USA, 1963; pp. 99–214.

(6) Owen, J.; Thornley, J. H. M. Covalent bonding and magnetic properties of transition metal ions. *Rep. Prog. Phys.* **1966**, *29*, 675−728.

(7) Al Ma'Mari, F.; Moorsom, T.; Teobaldi, G.; Deacon, W.; Prokscha, T.; Luetkens, H.; Lee, S.; Sterbinsky, G. E.; Arena, D. A.; MacLaren, D. A.; Flokstra, M.; Ali, M.; Wheeler, M. C.; Burnell G.; Hickey, B. J.; Cespedes, O. Beating the Stoner criterion using molecular interfaces. *Nature* **2015**, *524*, 69−74.

(8) Callsen, M.; Caciuc, V.; Kiselev, N.; Atodiresei, N.; Blügel, S. Magnetic Hardening Induced by Nonmagnetic Organic Molecules. *Phys. Rev. Lett.* **2013**, *111*, 106805/1–5.

(9) Atodiresei, N.; Brede, J.; Lazić, P.; Caciuc, V.; Hoffmann, G.; Wiesendanger, R.; Blügel, S. Design of the Local Spin Polarization at the Organic–Ferromagnetic Interface. *Phys. Rev. Lett.* **2010**, *105*, 066601/1–4.

(10) Friedrich, R.; Caciuc, V.; Atodiresei, N.; Blügel, S. Molecular Induced Skyhook Effect for Magnetic Interlayer Softening. *Phys. Rev. B* **2015**, *92*, 195407/1–5.

(11) Lin, M.-W.; Chen, P.-H.; Yu, L.-C.; Shiu, H.-W.; Lai, Y.-L.; Cheng, S.-L.; Wang, J.-H.; Wei, D.-H.; Lin, H.-J.; Chin, Y.-Y.; Hsu, Y.-J. Enhanced Magnetic Order and Reversed

Magnetization Induced by Strong Antiferromagnetic Coupling at Hybrid Ferromagnetic–Organic Heterojunctions. *ACS Appl. Mater. Interfaces* **2022**, *14*, 16901–16910.

(12) Carmeli, I.; Leitus, G.; Naaman, R.; Reich, S.; Vager Z. Magnetism induced by the organization of self-assembled monolayers. *J. Chem. Phys.* **2003**, *118*, 10372−10375.

(13) Carmeli, I.; Skakalova, V.; Naaman, R.; Vager Z. Magnetization of Chiral Monolayers of Polypeptide: A Possible Source of Magnetism in Some Biological Membranes. *Angew. Chem. Int. Ed.* **2002**, *41*, 761−763.

(14) Carmeli, I.; Leitus, G.; Naaman, R.; Reich, S.; Vager Z. New Electronic and Magnetic Properties of Monolayers of Thiols on Gold. *Isr. J. Chem.* **2003**, *43*, 399–405.

(15) Shen Y.; Chen, C.-F. Helicenes: Synthesis and Applications. *Chem. Rev.* **2012**, *112*, 1463–1535.

(16) Crassous, J.; Stará, I.; Stary, I. (eds.), Helicenes - Synthesis, Properties and Applications Wiley-VCH, Weinheim 2022.

(17) Bloom, B. P.; Paltiel, Y.; Naaman, R.; Waldeck, D. H. Chiral Induced Spin Selectivity. *Chem. Rev.* **2024**, *124*, 1950−1991.

(18) Naaman, R.; Paltiel, Y.; Waldeck, D. H. Chiral Molecules and the Electron Spin. *Nat. Rev. Chem.* **2019**, *3,* 250–260.

(19) Mori, T. Chiroptical Properties of Symmetric Double, Triple, and Multiple Helicenes, *Chem. Rev.* **2021**, *121*, 2373–2412.

(20) Gingras, M. One Hundred Years of Helicene Chemistry. Part 3: Applications and Properties of Carbohelicenes. *Chem. Soc. Rev.* **2013**, *42*, 1051–1095.

(21) Ernst, K.-H. Stereochemical recognition of helicenes on metal surfaces. *Acc. Chem. Res.* **2016**, *49*, 1182–1190.

(22) Ernst, K.-H. Helicenes on Surfaces: Stereospecific On-Surface Chemistry, Single Enantiomorphism, and Electron Spin Selectivity. *Chirality* **2024**, *36*, e23706/1–11.

(23) Safari, M. R.; Matthes, F.; Caciuc, V.; Atodiresei, N.; Schneider, C. M.; Ernst, K.-H.; Bürgler, D. Enantioselective adsorption on magnetic surfaces. *Adv. Mater.* **2024**, *36*, 2308666/1–8.

(24) Göhler, B.; Hamelbeck, V.; Markus, T. Z.; Kettner, M.; Hanne, G. F.; Vager, Z.; Naaman, R.; Zacharias, Z. *Science* **2011**, *331*, 894–897.

(25) V. Kiran, V.; Mathew, S. P.; Cohen, S. R.; Hernández Delgado, I.; Lacour, J; Naaman, R. Helicenes–A New Class of Organic Spin Filter. *Adv. Mater.* **2016**, *28*, 1957–1962.

(26) Mishra, D.; Markus, T. Z.; Naaman, R.; Kettner, M.; Göhler, B.; Zacharias, H.; Friedman, N.; Sheves, M.; Fontanesi, C. Spin-dependent electron transmission through bacteriorhodopsin embedded in purple membrane. *Proc. Natl. Acad. Sci. U.S.A.* **2013**, *110*, 14872–14876.

(27) Naaman, R.; Waldeck D. H. Chiral-induced spin selectivity effect. *J. Phys. Chem. Lett.* **2012**, *3*, 2178–2187.

(28) Safari, M. R.; Matthes, F.; Schneider, C. M.; Ernst, K.-H.; Bürgler, D. Spin-Selective Electron Transport Through Single Chiral Molecules. *Small* **2024**, *20*, 2308233/1–7.

(29) Banerjee-Ghosh, K.; Dor, O. B.; Tassinari, F.; Capua, E.; Yochelis, S.; Capua, A.; Yang, S.-H.; Parkin, S. S. P.; Sarkar, S.; Kronik, L.; Baczewski, L. T.; Naaman, R.; Paltiel, Y. Separation of enantiomers by their enantiospecific interaction with achiral magnetic substrates. *Science* **2018**, *360*, 1331−1334.

(30) Altman, M. S.; Pinkvos, H.; Hurst, J.; Poppa, H.; Marx, H.; Bauer, E. Spin Polarized Low Energy Electron Microscopy of Surface Magnetic Structure. *Mater. Res. Soc. Symp. Proc.* **1991**, *232*, 125−132.

(31) Pinkvos, H.; Poppa, H.; Bauer, E.; Hurst, J. Spin-polarized low-energy electron microscopy study of the magnetic microstructure of ultra-thin epitaxial cobalt films on W(110). *Ultramicroscopy* **1992**, *47*, 339−345.

(32) Telieps, W.; Bauer, E. An Analytical Reflection and Emission UHV Surface Electron Microscope. *Ultramicroscopy* **1985**, *17*, 57−66.

(33) Irziqat, B.; Berger, J.; Mendieta-Moreno, J. I.; Shyam Sundar, M.; Bedekar, A. V.; Ernst, K.-H. Transition from Homochiral Clusters to Racemate Monolayers During 2D Crystallization of Trioxa[11]helicene on Ag(100). *ChemPhysChem* **2021**, *22*, 293–296.

(34) Irziqat, B.; Berger, J.; Cebrat, A.; Mendieta-Moreno, J. I.; Shyam Sundar, M.; Bedekar, A. V.; Ernst, K.-H. Conglomerate Aggregation of 7,12,17-Trioxa[11]helicene into Homochiral Two-Dimensional Crystals on the Cu(100) Surface. *Helv. Chim. Acta* **2022**, *105*, e202200114/1−8.

(35) Baljozović, M.; Fernandes Cauduro, A. L.; Seibel, J.; Mairena, A.; Grass, S.; Lacour, J.; Schmid, A. K.; Ernst, K.-H. Growth Dynamics and Electron Reflectivity in Ultrathin Films of Chiral Heptahelicene on Metal (100) Surfaces Studied by Spin-Polarized Low Energy Electron Microscopy. *Phys. Status Solidi B* **2021**, *258*, 2100263/1−8.

(36) Rougemaille, N.; Schmid, A.; Magnetic imaging with spin-polarized low-energy electron microscopy. *Eur. Phys. J. Appl. Phys.* **2010**, *50*, 20101/1−18.

(37) Baljozović, M.; Fernandes Cauduro, A. L.; Seibel, J.; Mairena, A.; Grass, S.; Lacour, J.; Schmid, A. K.; Ernst, K.-H. Growth Dynamics and Electron Reflectivity in Ultrathin Films of Chiral Heptahelicene on Metal (100) Surfaces Studied by Spin-Polarized Low Energy Electron Microscopy. *Phys. Status Solidi B* **2021**, *258*, 2100263/1−8.

(38) Baljozović, M.; Arnoldi, B.; Grass, S.; Lacour, J.; Aeschlimann, M.; Stadtmüller, B.; Ernst, K.-H. Spin- and angle-resolved photoemission spectroscopy study of heptahelicene layers on Cu(111) surfaces. *J. Chem. Phys.* **2023**, *159*, 044701/1–8.

(39) Kettner, M.; Maslyuk, V. V.; Nürenberg, D.; Seibel, J.; Gutierrez, R.; Cuniberti, G.; Ernst, K.-H.; Zacharias, H. Chirality-dependent electron spin filtering by molecular monolayers of helicenes. *J. Phys. Chem. Lett.* **2018**, *9*, 2025–2030.

(40) Li, L.; Shi, W.; Mahajan, A.; Zhang, J.; Gómez-Gómez, M.; Labella, J.; Louie, S.; Torres, T.; Barlow, S.; Marder, S. R.; Reichman, D. R.; Venkataraman, L. Too Fast for Spin Flipping: Absence of Chirality-Induced Spin Selectivity in Coherent Electron Transport through Single-Molecule Junctions. *J. Am. Chem. Soc.* **2025**, *147*, 25043−25051.

(41) Zoppi, L.; Stöckl, Q.; Mairena, A.; Allemann, O.; Siegel, J. S.; Baldridge, K. K.; Ernst, K.-H. Pauli Repulsion Versus van der Waals: Interaction of Indenocorannulene with a Cu(111) Surface. *J. Phys. Chem. B* **2018**, *122*, 871−877.

(42) Mairena, A.; Zoppi, L.; Lampart, S.; Baldridge, K. K.; Siegel, J. S.; Ernst, K.-H. Fivefold Symmetry and 2D Crystallization: Self-Assembly of the Buckybowl Pentaindenocorannulene on a Cu(100) Surface. *Chem. Eur. J.* **2019**, *25*, 11555–11559.

(43) Anderson, P. W. Localized magnetic states in metals. *Phys. Rev.* **1961**, *124*, 41–53.

(44) Newns, D. M. Self-consistent model of hydrogen chemisorption. *Phys. Rev.* **1969**, *178*, 1123–1135.

(45) Şaşıoğlu, E.; Friedrich, C.; Blügel, S. Effective Coulomb Interaction in Transition Metals from Constrained Random-Phase Approximation. *Phys. Rev. B* **2011**, *83*, 121101/1–4.

(46) Shyam Sundar, M.; Bedekar, A. V. Synthesis and Study of 7, 12, 17-Trioxa[11] helicene. *Org. Lett.* **2015**, 17, 5808–5811.

(47) Blum, V.; Gehrke, R.; Hanke, F.; Havu, P.; Havu, V.; Ren, X.; Reuter, K.; Scheffler, M. Ab Initio Molecular Simulations with Numeric Atom-Centered Orbitals. *Comput. Phys. Commun.* **2009**, *180*, 2175−2196.

(48) Perdew, J. P.; Burke, K.; Ernzerhof, M. Generalized gradient approximation made simple. *Phys. Rev. Lett.* **1996**, *77*, 3865/1−4.

(49) Tkatchenko, A.; Scheffler, M. Accurate Molecular Van Der Waals Interactions from Ground-State Electron Density and Free-Atom Reference Data. *Phys. Rev. Lett.* **2009**, *102*, 073005/1−4.

(50) Adamo, C.; Barone, V. Toward Reliable Density Functional Methods without Adjustable Parameters: The PBE0Model. *J. Chem. Phys.* **1999**, *110*, 6158−6170.

(51) Fishman, M.; White, S. R.; Stoudenmire, E. M. The ITensor Software Library for Tensor Network Calculations. *SciPost Phys. Codebases* **2022**, *4*, 1−53. doi:10.21468/SciPostPhysCodeb.4

TOC graphics

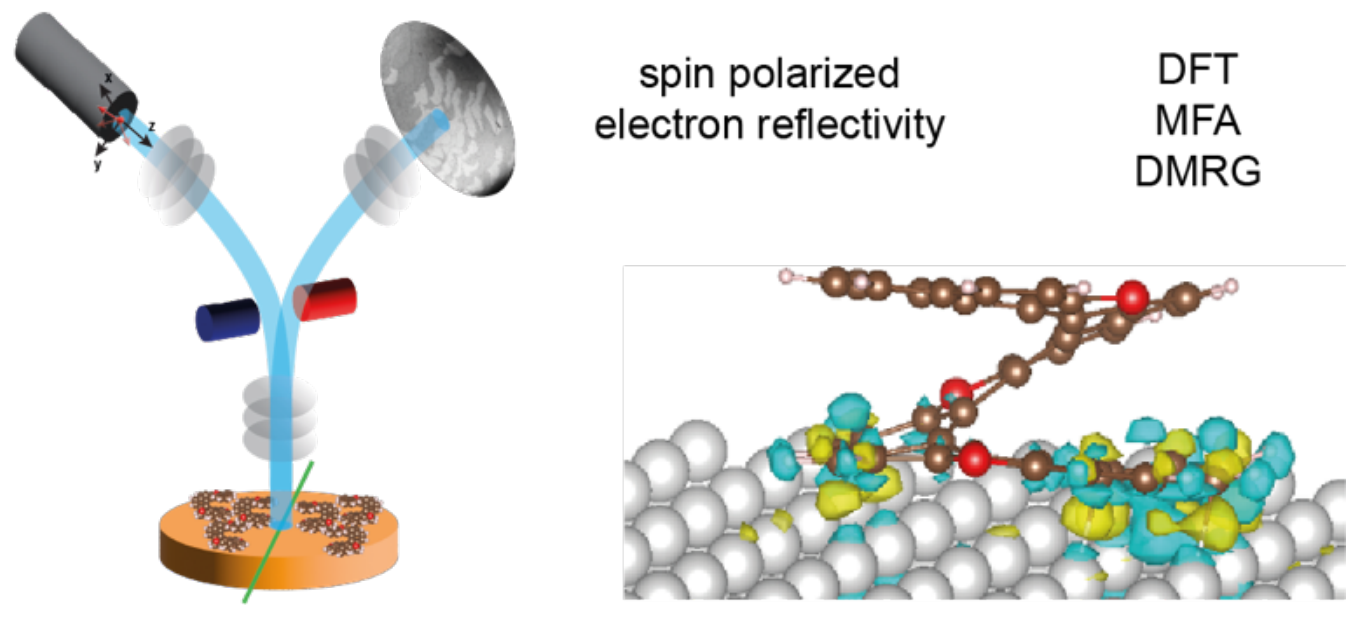

Supporting Information

# Adsorption-induced surface magnetism

Miloš Baljozović,[1] Shiladitya Karmakar,[2] André L. Fernandes Cauduro,[3] Mothuku Shyam Sundar,[4] Marco Lozano,[2] Manish Kumar,[2] Diego Soler Polo,[2] Andreas K. Schmid,[3] Ashutosh V. Bedekar,[4] Pavel Jelinek[2]* & Karl-Heinz Ernst[1,2]*

[1] Empa, Swiss Federal Laboratories for Materials Science and Technology; Dübendorf, Switzerland.

[2] Nanosurf Lab, Institute of Physics of the Czech Academy of Sciences; Prague, Czech Republic.

[3] National Center for Electron Microscopy, Molecular Foundry, Lawrence Berkeley National Laboratory; Berkeley, CA 94720, USA

[4] Department of Chemistry, The Maharaja Sayajirao University of Baroda; Vadodara 390 002, India

e-mail: jelinekp@fzu.cz ; kalle@fzu.cz

## SI Figures

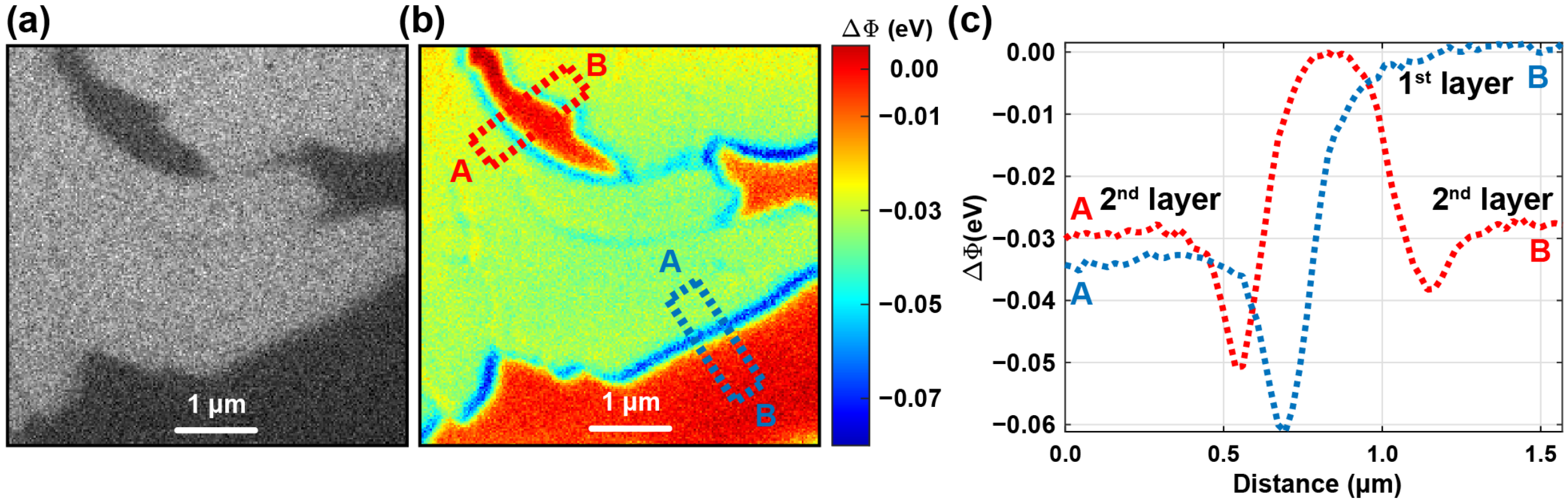

**Figure S1.** Work function difference between 1st and 2nd layer of *(M)*-TO[11]H. a) LEEM bright field image acquired using start voltage of 3.5 V where first layer appears dark with second layer appearing bright. b) Pixel-by-pixel map of ΔΦ of the film shown in (a) plotted in colour scale shown on the right side of the panel, representing work function difference between the two layers. c) line profiles from areas marked in b) showing that the double layer regions have about 30 meV lower work function with respect to the 1 ML regions. Work-function changes at the boundaries between layers are artifacts originating from the beam drift during the measurements.

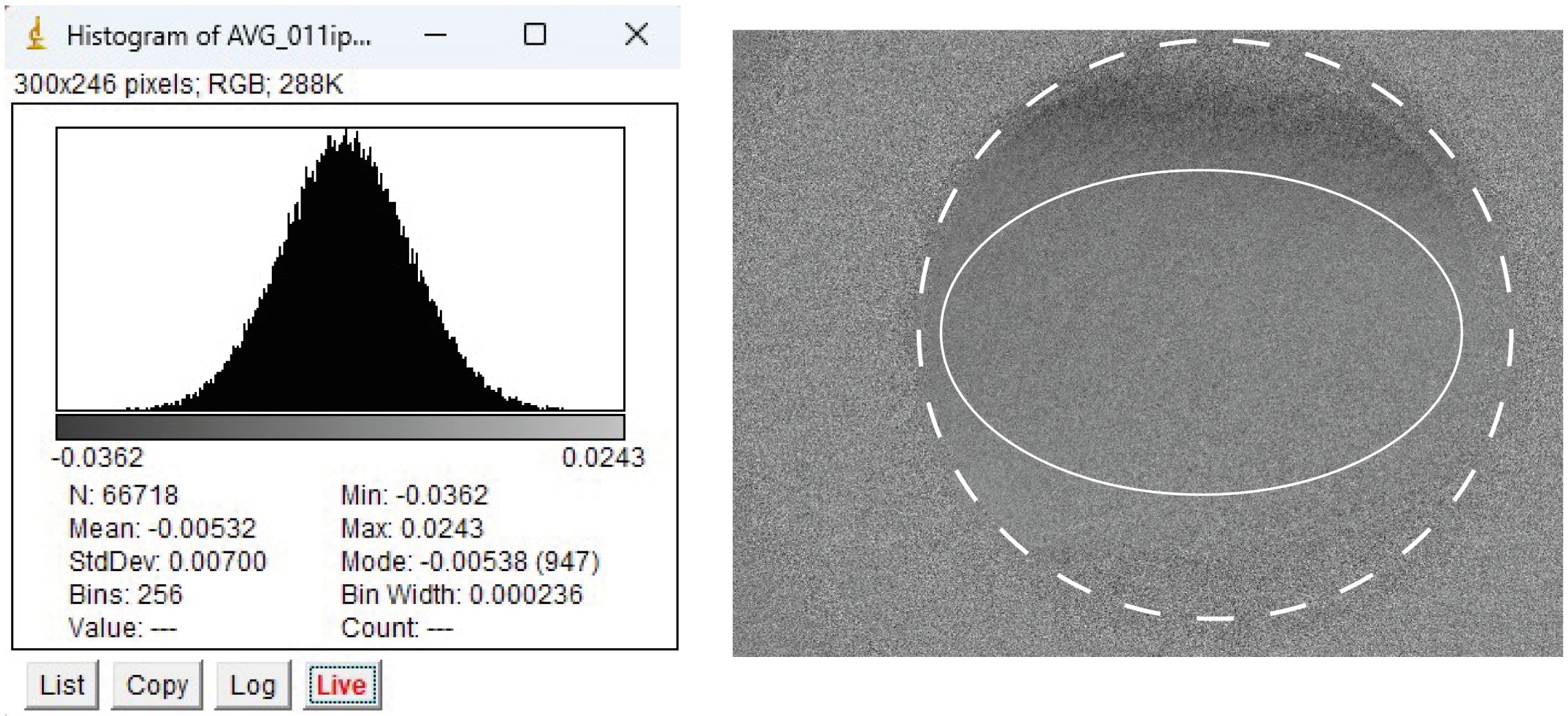

**Figure S2.** Example spin asymmetry image calculated according to Equation 1 (right). Dashed circle highlights the complete LEEM field of view, while ellipse denotes area in the field of view from which the average spin-asymmetry values were determined. The areas from field of view on top and bottom ends with less homogeneous illumination are discarded. Histogram (left) of spin asymmetry values from the area encircled with the ellipse. The distribution of the values is used for estimating errors of the spin asymmetry values.

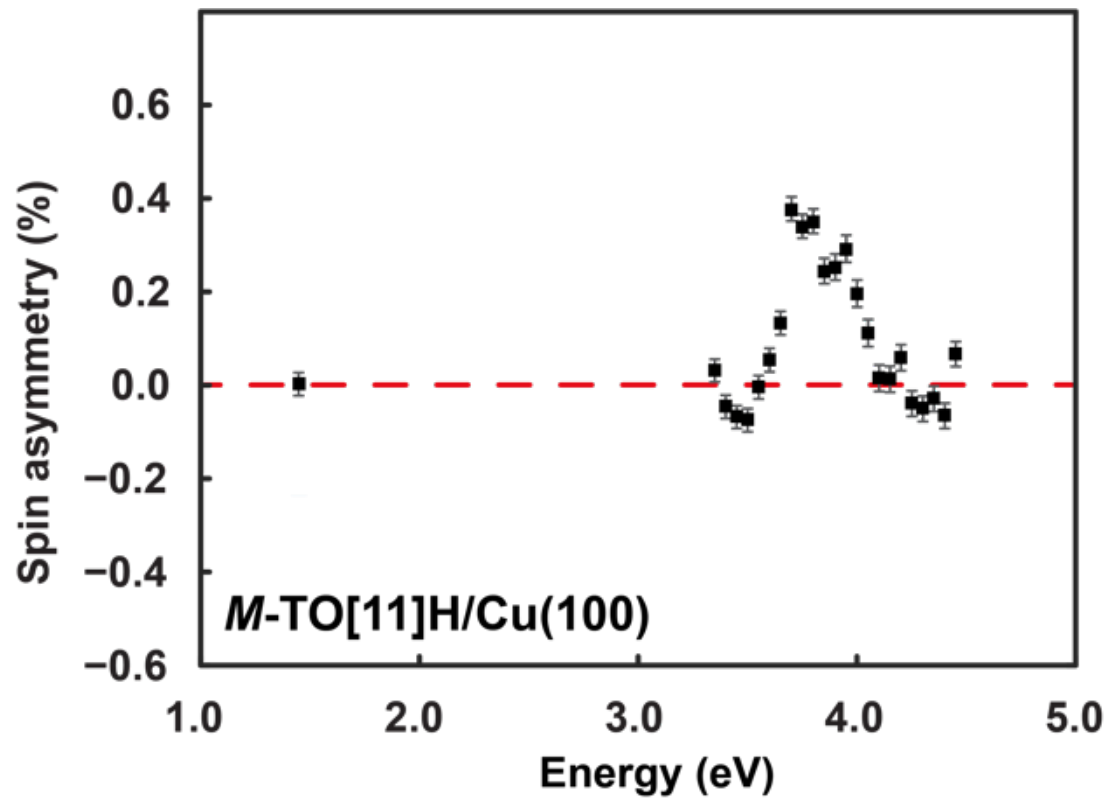

**Figure S3.** By inversion of the spin-polarization sequence a reversal of asymmetry with respect to the negative dips shown in Figures 2 and S4.

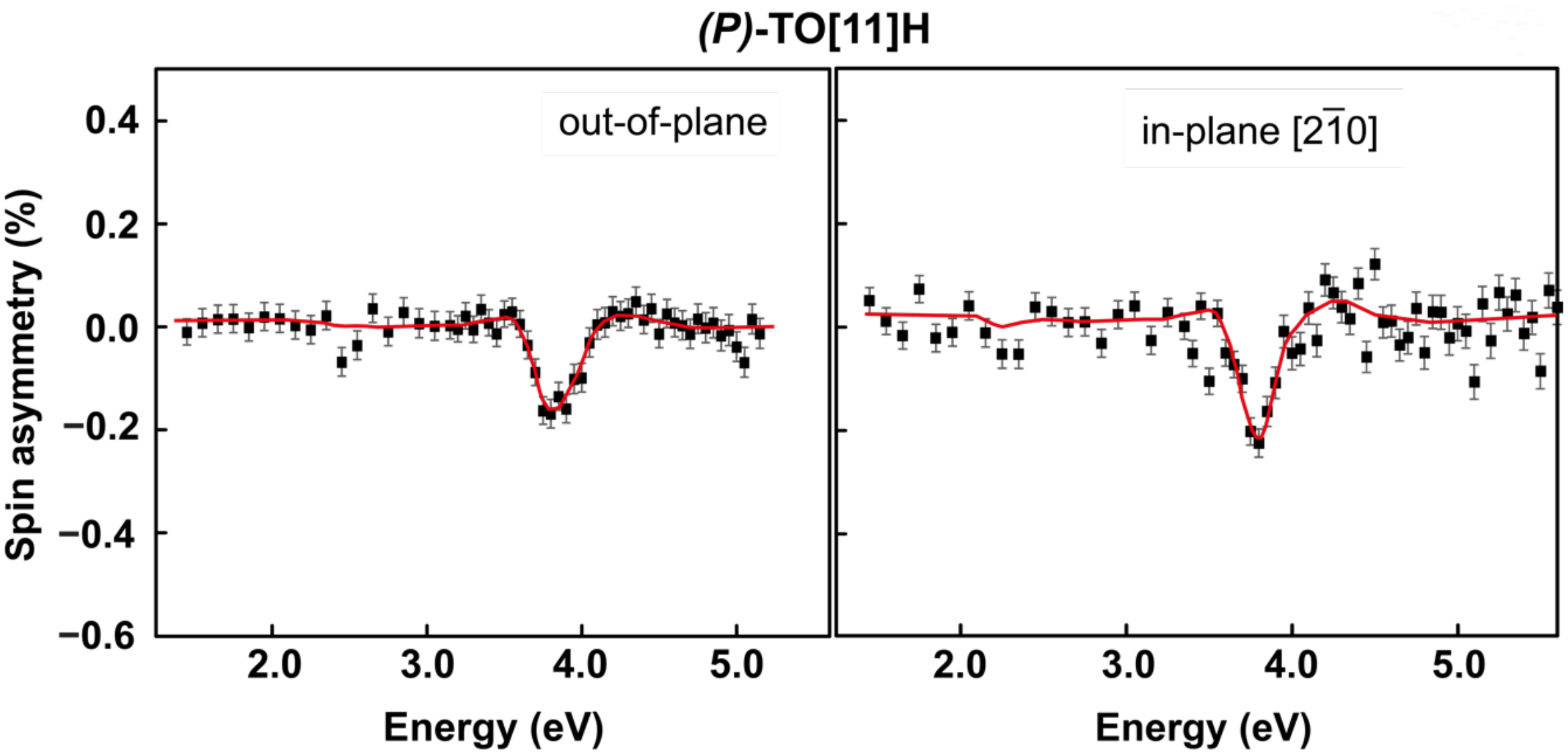

**Figure S4.** Spin asymmetry plots of *(P)*-TO[11]H on Cu(100). Out-of-plane (left) and in-plane [2$\bar{1}$0] (right) spin asymmetry plots of *(P)*-TO[11]H on Cu(100). No difference compared to the asymmetry plots of *(M)*-TO[11]H on Cu(100) shown in Figure 2 is observed which excludes chirality-induced spin selectivity as origin.

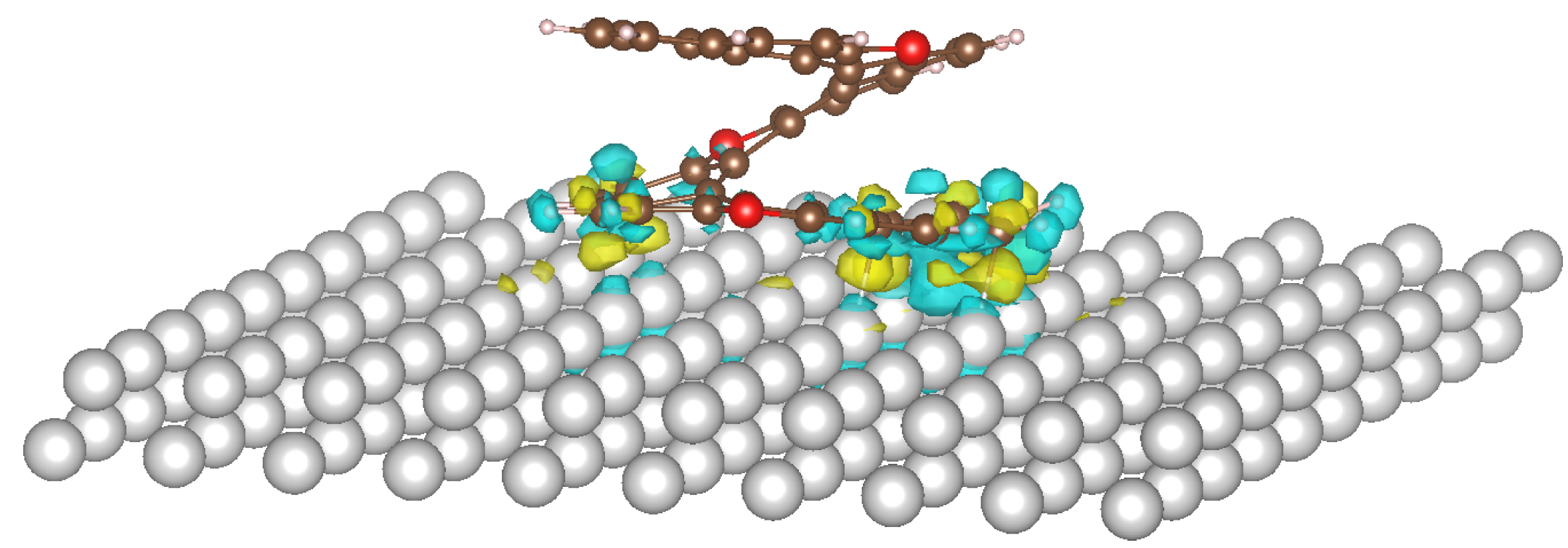

**Figure S5.** Sketch of the charge density difference Δρ of model system containing the TO[11]H molecule placed on 10×10 Cu(100) slab calculated using the PBE exchange–correlation functional. The isosurface is set at 0.02 eV/Å$^3$. Charge depletion and accumulation are shown as cyan and yellow isosurfaces, respectively, with $\Delta\rho > 0$ indicating accumulation and $\Delta\rho < 0$ indicating depletion.

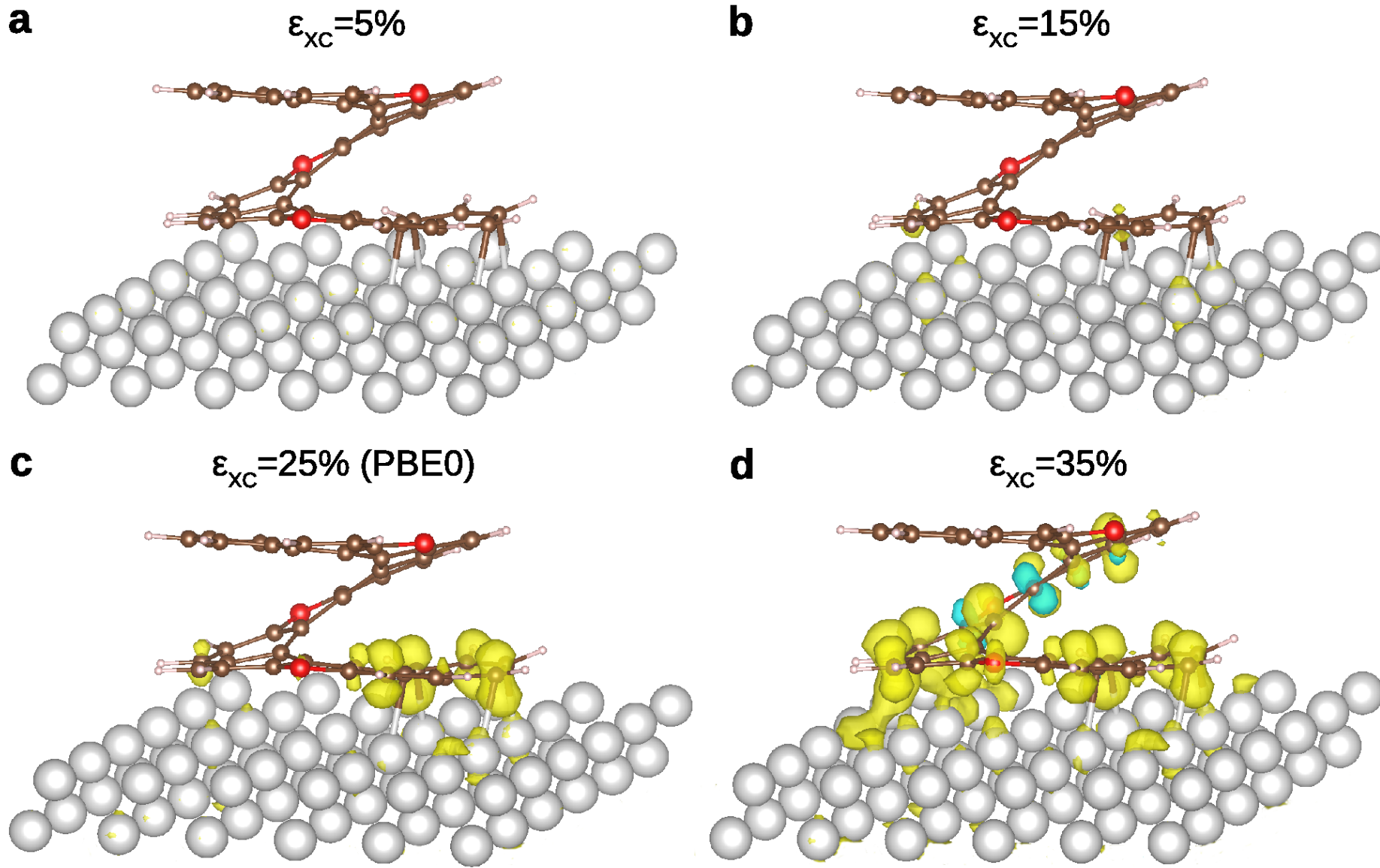

**Figure S6.** Sketch of the spin density of model system by varying the non-local Hartree Fock exchange ($\varepsilon_{xc}$) that contributes to the hybrid functional PBE0. Spin density accumulation is shown for a) $\varepsilon_{xc}$=5%, b) $\varepsilon_{xc}$=15%, c) $\varepsilon_{xc}$=25%, and d) $\varepsilon_{xc}$=35%. The isosurface is set at 0.02 e/Å$^3$ with accumulated spin is shown by yellow colored isosurface. Spin density increases as $\varepsilon_{xc}$ is increased. $\varepsilon_{xc}$ = 25% is the case of standard PBE0.

**Phase diagram in the ($U_d$, $t_{dH}$) and ($U_d$, $t_{sH}$) parameter spaces:**

Figure S7 shows phase diagrams of $M_{HOMO}$ in the ($U_d$, $t_{dH}$) and ($U_d$, $t_{sH}$) planes. Two regions are observed: a non-magnetic phase and a magnetic phase with finite $M_{HOMO}$. In panel (a), the phase boundary shifts to larger $U_d$ as $t_{dH}$ increases. Suppression of the magnetic solution for large values of $t_{dH}$ is observed. The magnetic phase boundary in the $t_{dH}$-$U_d$ plane is nearly linear, with slope ~0.3. This indicates that a magnetic solution emerges only when $U_d/t_{dH}$ ~ 3.33. In panel (b), increasing $t_{sH}$ suppresses magnetism, expanding the non-magnetic region. But relatively weak dependence of the $U_d$ threshold on $t_{sH}$ is observed.

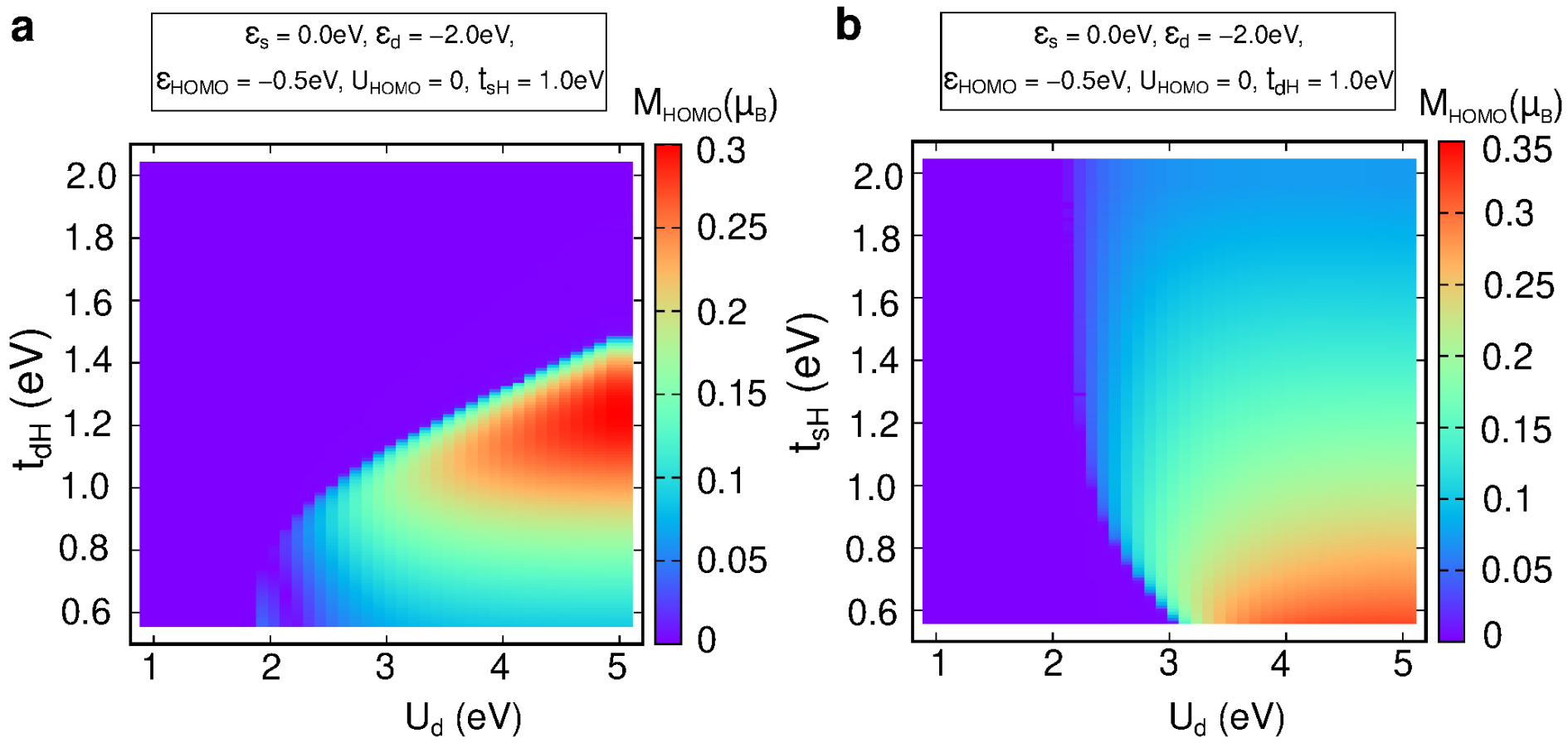

**Figure S7.** (a) $t_{dH}$-$U_d$ phase diagram with $t_{sH}$ = 1.0 eV. (b) $t_{sH}$-$U_d$ phase diagram with $t_{dH}$ = 1.0 eV. In both panels, $\varepsilon_s$ = 0 eV, $\varepsilon_d$ = −2.0 eV, $\varepsilon_{HOMO}$ = −0.5 eV, and $U_{HOMO}$ = 0. Two distinct regions are observed depending of $M_{HOMO}$: a) non-magnetic region, marked in indigo, b) magnetic region, where $M_{HOMO} < 0.25\mu_B$, marked in green and $M_{HOMO} > 0.25\mu_B$, marked in red.

**Phase diagram in the ($U_d$, $U_{HOMO}$) and ($U_d$, $\varepsilon_{HOMO}$) parameter spaces:**

Figure S8 shows phase diagrams of $M_{HOMO}$ as a function of $U_d$ and the molecular parameters $U_{HOMO}$ (panel a) and $\varepsilon_{HOMO}$ (panel b). Two regions are observed similar to Figure S7. In panel (a), a sharp phase boundary is observed, indicating that the threshold value of $U_d$ is insensitive in $U_{HOMO}$, thus correlation on the HOMO plays only a minor role. In contrast, panel (b) shows that $\varepsilon_{HOMO}$ strongly affects the onset of magnetism: lowering $\varepsilon_{HOMO}$ stabilizes spin polarization by reducing the $U_d$ threshold. Notably, magnetism is favored only when the HOMO lies no more than ~ 1.5 eV below $E_F$. To benchmark the mean-field results, we also calculated the occupations of natural orbitals by solving the model Hamiltonian using DMRG with the same model parameters. From these occupations of natural orbitals, the radical character was evaluated following the method proposed by Yamaguchi et al.,[1] which provides insight into the magnetization of the molecule (Figure S9).

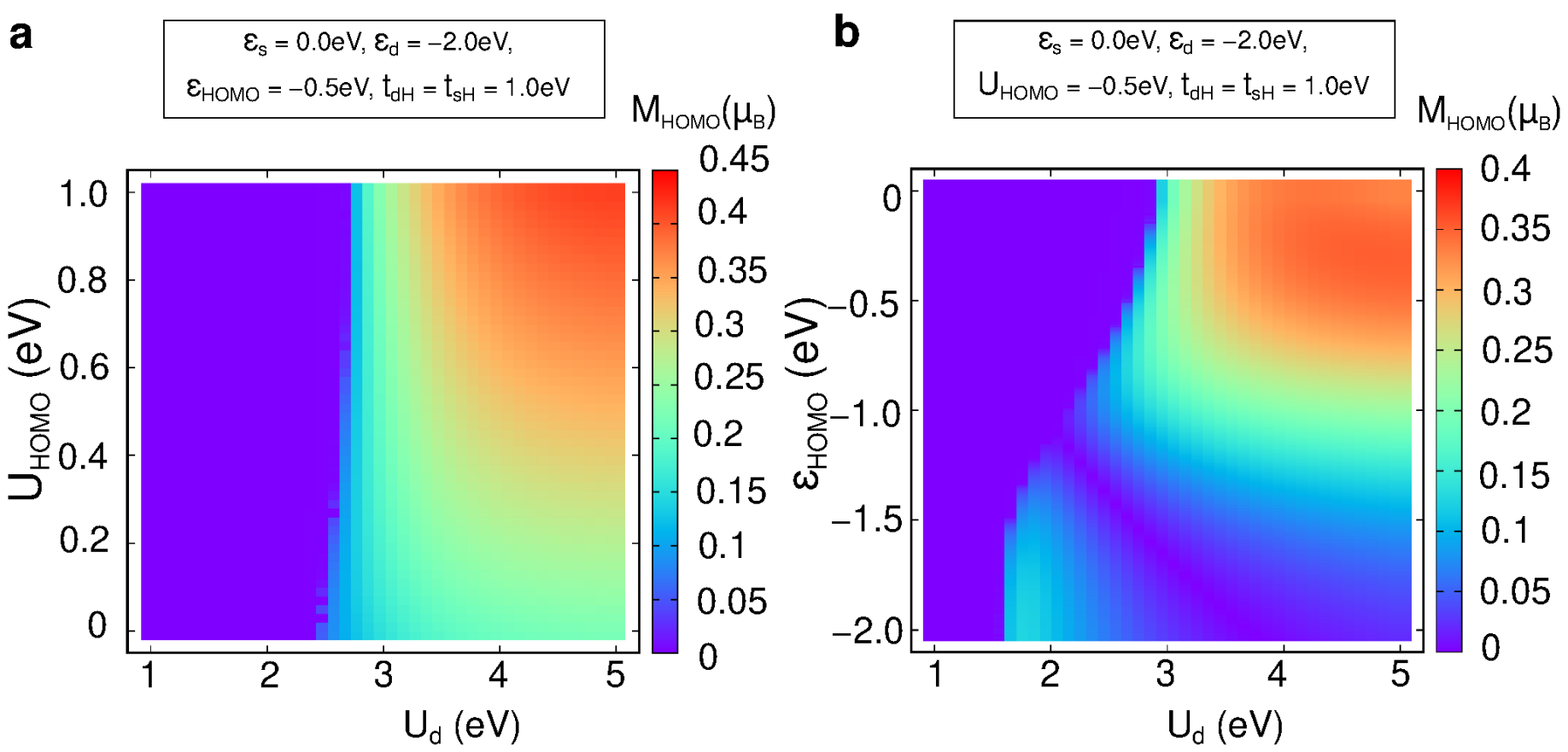

**Figure S8.** (a) $U_{HOMO}$ vs $U_d$ phase diagram with $\varepsilon_{HOMO}$ set at –0.5 eV showing distinct phases depending of $M_{HOMO}$. (b) $\varepsilon_{HOMO}$ *versus* $U_d$ phase diagram with $U_{HOMO}$ set at 0.5 eV. Different regions can be marked as following: i) non-magnetic region (marked in indigo), ii) magnetic region, where $M_{HOMO} < 0.25\mu_B$ (marked in green), and $M_{HOMO} > 0.25\mu_B$ (marked in red).

(1) Yamaguchi, K.; Takahara, Y.; Fueno, T.; Houk, K. N. Extended Hartree-Fock (EHF) theory of chemical reactions. III. Projected Møller-Plesset (PMP) perturbation wavefunctions for transition structures of organic reactions. *Theoret. Chim. Acta* **1988,** *73*, 337–364.

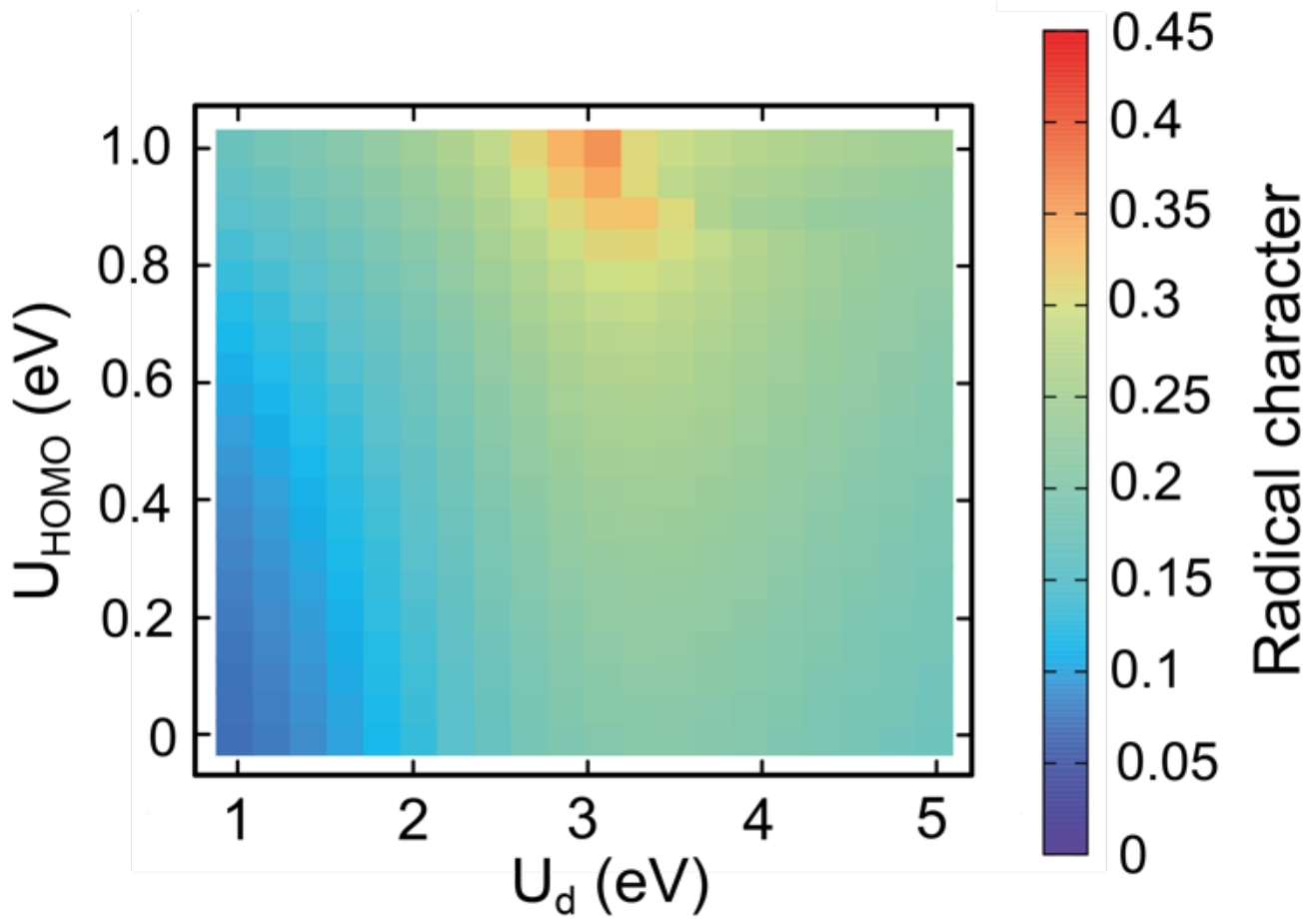

**Figure S9.** Phase diagram of $U_{HOMO}$ vs $U_d$, obtained using the same parameters as in Figure S8a, showing the radical character of the molecule. The radical character is calculated using the method proposed by Yamaguchi et al.,[1] based on the occupation number of natural orbitals obtained from the DMRG calculations.

**Chiral separation of trioxa[11]helicene on preparative HPLC**

The resolution of racemic trioxa[11]helicene into its enantiomers was performed by HPLC using a Chiralart Cellulose SC column (250 x 4.6 mm, 5 um, YMC) and *n*-heptane/2-propanol (95:05) mixture as the mobile phase. The racemic sample is dissolved in *n*-heptane/2-propanol, injected on the chiral column, and detected with an UV detector at 254 nm. The earlier eluting fractions gave the enantiomer exhibiting a positive optical rotation (+) which was isolated in 40% yield and 100% ee. Later eluting fractions consisted of the enantiomer exhibiting a negative optical rotation (-) in 42% yield with > 95% ee. The enantiomeric purity of both enantiomers was checked by chiral HPLC using the same stationary phase. The optical rotations were measured in chloroform using an Autopol IV instrument (Rudolph Research Analytical) and the specific rotation values obtained for (+)-**E1** (*P*-isomer) and (–)-**E2** (*M*-isomer) were found to be $[\alpha]^{25}_{D}$ +3803 ($c$ = 0.160, $CHCl_3$) and –3610 ($c$ = 0.154, $CHCl_3$), respectively. The very high specific rotation value is a known feature of helical structures.

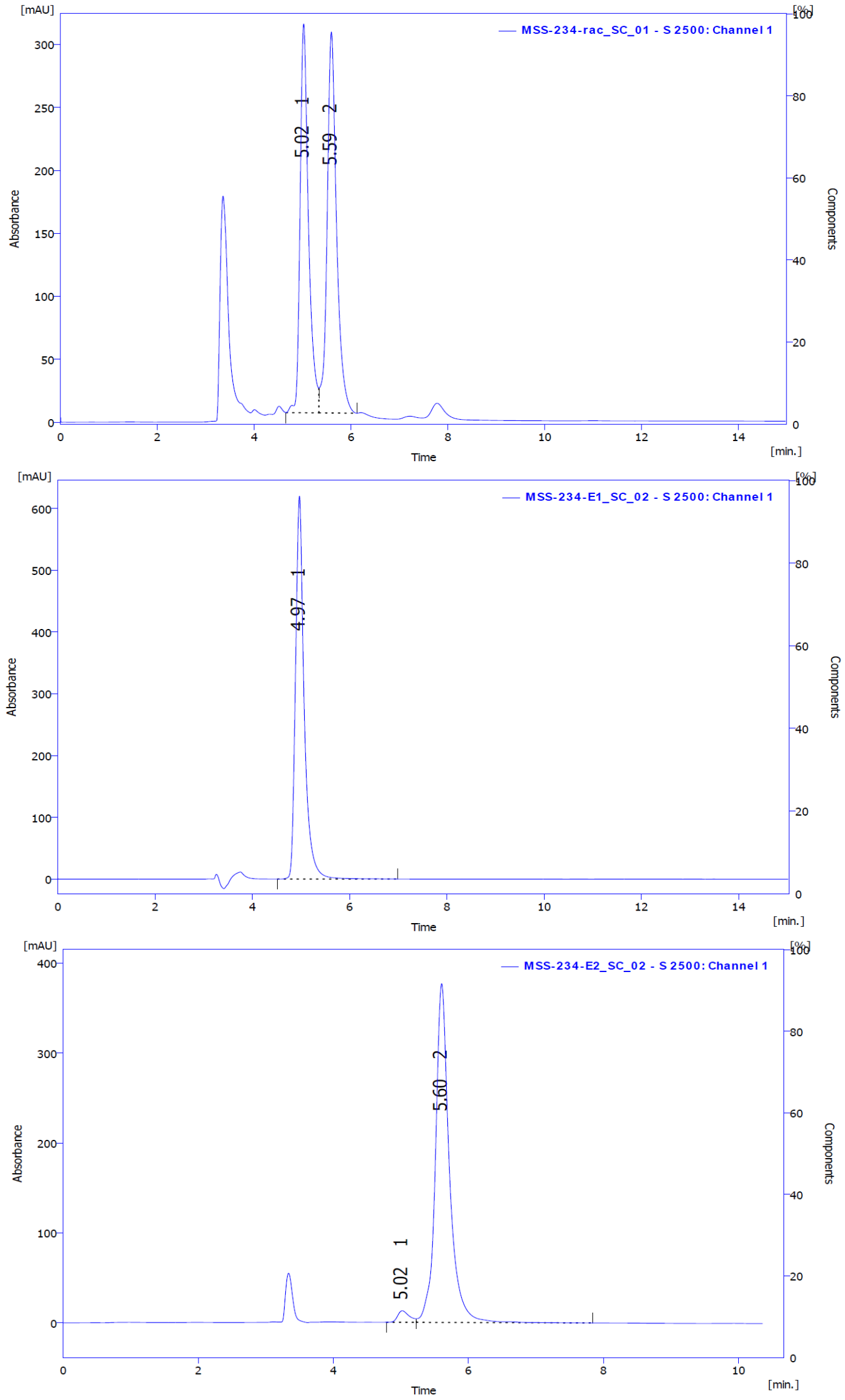

**Figure S10.** HPL chromatograms of racemic TO[11]H (top) and of the two separated enantiomers (middle, bottom). Column: Cellulose SC (250 x 4.6 mm, 5 um, YMC), Mobile Phase: heptane - i-PrO H 95:5, Flow Rate: 1.0 mL/min., Detection: UV.

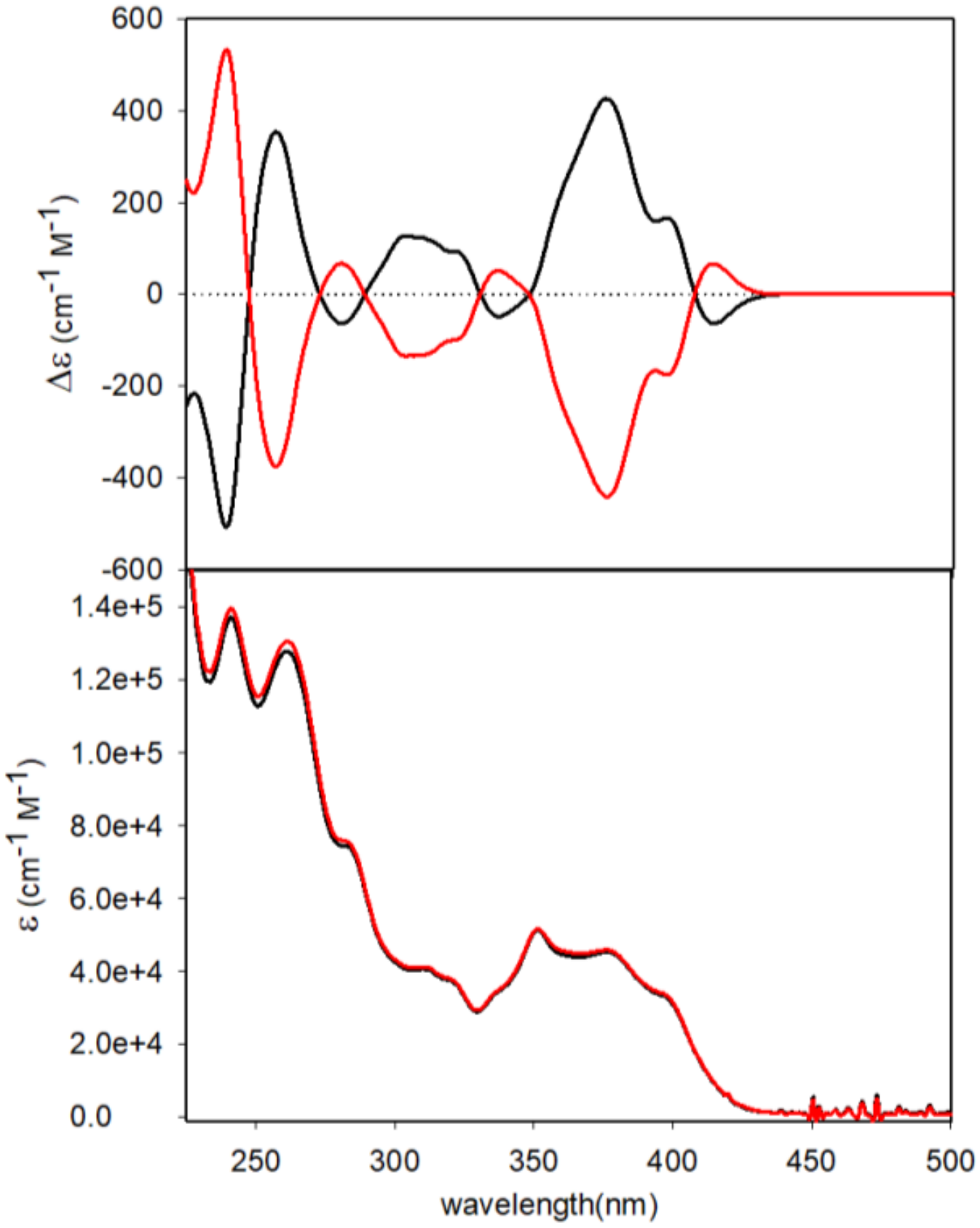

**Figure S11.** Circular dichroism (upper curve), red line (–)-(*M*)-**E2** and black line (+)-(*P*)-**E1** and UV−vis-absorption (lower curve) of trioxa[11]helicene (concentration = 0.1 mM in THF, 25 °C).

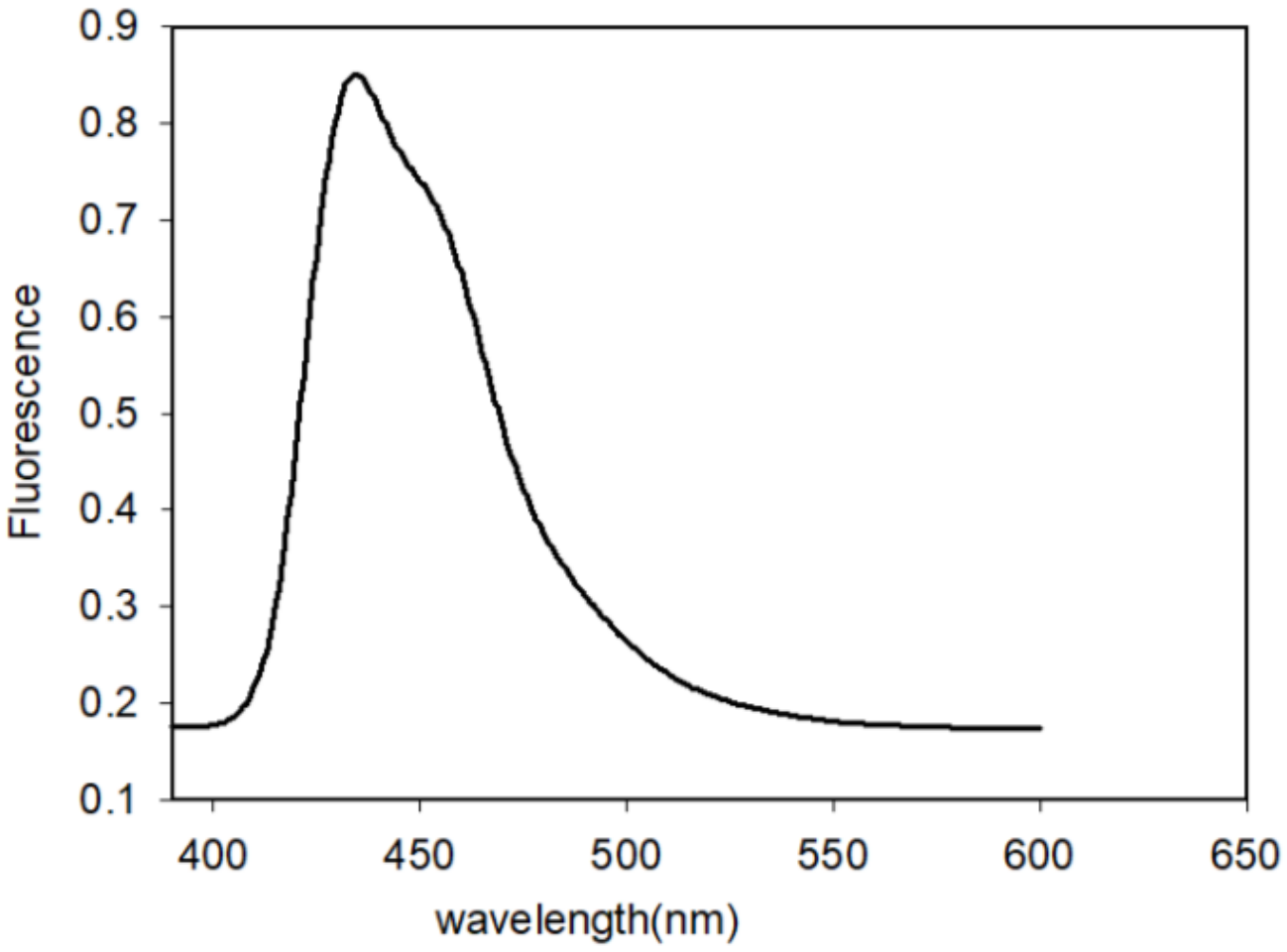

**Figure S12.** Fluorescence spectrum of trioxa[11]helicene (concentration = 0.1 mM in THF, 25 °C) upon excitation at 395 nm.